\documentclass[aps,prd,showkeys,showpacs,showkeys,nofootinbib]{revtex4}

\usepackage{amssymb}
\usepackage{amsmath}
\usepackage[dvipsone]{graphics}
\usepackage{psfig}
\usepackage{epsfig}
\usepackage{bm}
\expandafter\ifx\csname package@font\endcsname\relax\else
\expandafter\expandafter
 \expandafter\usepackage
 \expandafter\expandafter
 \expandafter{\csname package@font\endcsname}%
\fi

\begin{document}
\title{Aberration and the Fundamental Speed of Gravity in the Jovian Deflection Experiment}
\author{Sergei M. Kopeikin}
\email{kopeikins@missouri.edu}
\affiliation{Department of Physics \& Astronomy, University of
Missouri-Columbia, Columbia, MO 65211, USA}
\author{Edward B. Fomalont}
\email{efomalon@nrao.edu}
\affiliation{National Radio Astronomy Observatory, Charlottesville, VA 
22903, USA}
\begin{abstract}
We describe our explicit Lorentz-invariant solution of the Einstein and null geodesic
equations for the deflection experiment of 2002 September 8 when a
massive moving body, Jupiter, passed within 3.7' of a line-of-sight to a distant quasar.  We
develop a general relativistic framework which shows that our measurement of
the retarded position of a moving light-ray deflecting body (Jupiter) by making use of the gravitational time delay of quasar's radio wave is equivalent to comparison of the relativistic laws of the Lorentz transformation for gravity and light. Because, according to Einstein, the Lorentz transformation of gravity field variables must depend on a fundamental speed $c$, its measurement through the retarded position of Jupiter in the gravitational time delay allows us to study the causal nature of gravity and to set an upper limit on the speed of propagation of gravity in the near zone of the solar system as contrasted to the speed of the radio waves. 
 In particular, the $v/c$ term beyond of the standard Einstein's deflection, which we
measured to 20\% accuracy, is associated with the aberration of the null direction of the
the gravity force (``aberration of gravity'') caused by 
the Lorentz transformation of the Christoffel symbols from the static frame of Jupiter to the moving frame of observer. General relativistic formulation of the experiment
identifies the aberration of gravity with the retardation of gravity because the speed of gravitational waves in Einstein's theory is equal to the speed of propagation of the gravity force.
We discuss the misconceptions which have inhibited the acceptance of
this interpretation of the experiment.  We also comment on other
interpretations of this experiment by Asada, Will, Samuel, Pascual-S\'anchez, and Carlip and show
that their ``speed of light" interpretations confuse the Lorentz transformation for gravity with that for light, and the fundamental speed of gravity 
with the physical speed of light from the quasar. For this reason, the ``speed of light" interpretations
are not entirely
consistent with a retarded Li\'enard-Wiechert solution of the Einstein
equations, and do not properly incorporate how the phase of the radio
waves from the quasar is perturbed by the retarded gravitational field
of Jupiter.  Although all of the formulations predict the same
deflection to the order of $v/c$, our formulation shows that the
underlying cause of this deflection term is associated with the aberration of gravity and not of light, and
that the interpretations predict different deflections at higher
orders of $v/c$ beyond the Shapiro delay, thus, making their measurement highly desirable for deeper testing of general relativity in future astrometric experiments like Gaia, SIM, and SKA.
\vspace{1cm}
\end{abstract}
\maketitle
\clearpage
\tableofcontents
\clearpage

\section{Introduction}\label{intro}

The recent Very Long Baseline Interferometric (VLBI) experiment of the
ultra-precise measurement of the relativistic deflection of a quasar's
radio waves (light) by the gravitational field of Jupiter
\cite{kop2001,fk-apj} has captivated the keen interest of relativists
who were trying to deeply understand gravitational physics.  The
experiment and its physical interpretation has illustrated some of the important
subtleties in the understanding, the approaches, and the mathematical
techniques of general relativity (GR).

   The experimental goal was the measurement of the (non-radial or tangential) deflection
component associated with the retarded position of Jupiter caused by
its orbital motion and mathematically connected to the finite value of the fundamental speed entering in front of the time derivatives from the metric tensor in the
gravity field and null geodesic equations (that is, the fundamental constant $c$ in the Christoffel symbols and the curvature tensor). This fundamental speed of gravity causes the retardation in propagation of gravity force from a moving massive object which can be measured in the light-ray deflection experiments. In this paper we use general theory of relativity and denote the fundamental speed of gravity by $c$. Label $c_g$ for the speed of gravity is understood as a parameter in the experimental fitting of observations to the Einstein theory. The symbol $c$ is frequently called as the speed of light. However, one must clearly separate the relativistic effects caused by the physical speed of light from those associated with the fundamental speed of gravity in the Eiunstein equations. 
We shall show in the present paper how to distinguish these effects to avoid ambiguity. 
 
Einstein's general principle of relativity assumes that the gravity field obeys to the special principle of relativity and transforms from one inertial frame to another in accordance with the Lorentz transformation of the Minkowski space-time. Hence, it postulates that in a geometrized system of units \cite{mtw} the speed of gravity $c=1$ in any frame of reference. 
We measured the impact of the retarded gravitational field of Jupiter on light
deflection (as well as the instantaneous Shapiro deflection of $1100~\mu$arcsec) of
$50\pm 9~\mu$arcsec, in excellent agreement with the GR prediction of
$51~\mu$arcsec based on assumption that $c=1$ irrespectively of the reference frame.  We have interpreted
these result as a measurement of the fundamental speed of gravity $c=1.06\pm 0.21$ and confirmation that gravity interacts with moving particles (photons) not instantaneously.   

   However, at present there are four other interpretations, which do
not agree among themselves, concerning the results of the experiment
\footnote{Recent publication of Frittelli \cite{frit1} agrees
with our understanding of the gravitational physics of the Jovian
deflection experiment}.
\begin{itemize}
\item the experiment measured the speed of the radio waves from the
quasar in the static gravitational field of Jupiter \cite{ass},
\item the experiment measured the aberration of the radio waves from
the quasar \cite{will-apj,carlip},
\item the experiment measured no significant property of light or of
gravity \cite{samuel},
\item the experiment measured the light-time delay discovered by R{\o}mer in 1676 \cite{pask} 
\end{itemize}
\vskip 0.1cm\noindent 

In this paper, we streamline our previous theoretical calculations by
working out an explicit Lorentz-invariant Li\'enard-Wiechert solution
of the Einstein gravity field equations in \S 2.  The VLBI measurement of the phase
of a radio wave, which has propagated in the field of a moving
body (Jupiter), is described in \S 3.  We demonstrate that the fundamental speed
of the characteristics of the gravity field equations (which we call
for brevity ``the speed of gravity") and the aberration of gravity force are
indeed associated with the experiment.  The major misconceptions about
the experiment and properties of gravity in general relativity are
summarized in \S 4.  Finally, in \S 5 we outline why these alternative
interpretations of the experimental results differ with our general
relativistic interpretation, even though all interpretations predict
the same angular deflection result to order $v/c$ beyond the Shapiro delay. Appendix describes how light propagates in a simple bi-metric model of gravity with two metrics, and provides a clear evidence that a light-ray deflecting body deflects light from its retarded position connected to observer by a null line lying on the gravity null cone that is because gravity propagates with finite speed.

\section{The Metric Tensor, Retarded Time and the Fundamental Speed}

\subsection{Formulation} 

      Our interpretation of the experiment is based solely on General
Relativity and the physical meaning of the retarded Li\'enard-Wiechert
potentials used to solve the Einstein equations. These potentials
describe the propagation of the gravity field even in the case of
gravitating bodies which move uniformly with constant speed
\cite{carl} exactly as in electrodynamics where the retarded
Li\'enard-Wiechert potentials describe electromagnetic field of a
moving point charge \cite{jack, fe}.  We emphasize that although
gravitational waves, which amplitude is falling off as $1/r$, are not generated by a uniformly moving body the
null characteristics of the gravitational field are precisely defined
by the Li\'enard-Wiechert potentials through their property of the Lorentz invariance and the principle of physical causality.  Thus, our use of the ``speed of
gravity" term is a synonym for the ultimate speed of the null characteristics
of the gravity field equations.  This speed is numerically equal to
the fundamental speed of the Minkowski geometry which is associated with the speed of light in vacuum. However, in general relativity the fundamental speed could be different from the speed of light. In the near-field zone of a gravitating system the null characteristics of gravity reveal themselves through the Lorentz invariance of the gravitational force (the aberration of gravity) when one compares observations done in one inertial frame with another, while in the far-field radiative zone the null characteristics of gravity can be traced as freely propagating gravitational waves. 

In general relativity light and gravity null rays are bi-characteristics, that is they propagate with the same fundamental speed. At the first glance it may look that they are identical and can not be discriminated. However, this point of view is erroneous. Although light and gravity propagate with the same speed on the hyper-surface of one and the same null cone, they (1) generally propagate in different directions in space in each particular gravitational experiment, and (2) they are associated with different physical effects (the aberration of light and the aberration of gravity, for instance). Hence, discrimination between the light and gravity null cones is possible if one makes use of additional properties of the propagating system of the light and gravity null rays.  

   Our definition of the ``speed of gravity" is more general than that
used by Asada, Samuel, and Will \cite{ass,will-apj, samuel} who
limited its meaning as the speed of propagation of free gravitational
waves alone.  In their formulations of the experiment, these authors assumed
only far-field gravitational effects, where gravitational waves are
dominant and, hence, the differentiation between the speed of gravity $c$ and that of light occurs only at
orders of $(v/c)^2$ beyond the Shapiro delay and higher.  This was one reason why the ``speed
of light'' was interpreted as causing the observed aberration of gravity force of Jupiter.
However, the experiment was performed in the near-field of the quasar
radio wave-Jupiter interaction where the transformational degrees of freedom for gravity not
associated with gravitational waves are dominant but they must not be confused with the transformational degrees of freedom for radio waves used for observations.

It is clear that there are different theoretical frameworks to
interpret the Jovian deflection experiment as a new test of
non-stationary properties of the gravitational field in the solar
system.  However, our point of view is that the most natural
interpretation should rest solidly on GR which has passed all other
experimental tests in strong compliance with our current theoretical
understanding of gravitational physics. In other words, in order to
interpret the experiment properly, the space-time properties which GR
postulates must not be violated; that space-time is a differentiable
manifold endowed with an affine connection (Christoffel symbols) whose
geodesics form a privileged set of world-lines in space-time, and
their knowledge (observation) allows us to extract information about
the curvature tensor (tidal gravitational force) without ambiguity.  The Christoffel
symbols define the law of motion of test particles and extended bodies and, hence, generalize the concept of the Newtonian gravitational force in GR. They
are formed from partial derivatives of the metric tensor that
defines geometric properties of the space-time and the causal structure
of the null cone. At the same time the metric tensor is associated
with a gravitational potential whose properties are determined by
Einstein's equations. 

In what follows the Greek indices $\alpha, \beta,...$ run from 0 to 3,
the Roman indices $i,j,...$ run from 1 to 3, repeated Greek indices
mean Einstein's summation from 0 to 3, and bold letters ${\bm
a}=(a^1,a^2,a^3), {\bm b}=(b^1,b^2,b^3),$ etc. denote spatial
(3-dimensional) vectors. A dot between two spatial vectors, for example
${\bm a}\cdot{\bm b}=a^1b^1+a^2b^2+a^3b^3$, means the Euclidean dot
product, and the cross between two vectors, for example ${\bm
a}\times{\bm b}$, means the Euclidean cross product. We also use a
shorthand notation for partial derivatives $\partial_\alpha
=\partial/\partial x^\alpha$. Greek indices are raised and lowered with 
full metric $g_{\alpha\beta}$. The Minkowski (flat) space-time metric $\eta_{\alpha\beta}={\rm diag}(-1,+1,+1,+1)$.

The solution of the Einstein equations, using the Li\'enard-Wiechert
potentials has been essentially understood and developed by previous researchers and we do not intend to go into complete details of this iterative procedure called by the post-Minkowskian approximations \cite{damour83}. For our purposes the first (linearized gravity) post-Minkowskian approximation is sufficient since the gravitational light-ray deflection experiments are not yet sensitive enough to measure non-linearities of the Einstein equations. The linearized gravity approximation is as follows.

We introduce the weak-field decomposition of
the metric tensor
\begin{equation}
\label{y}
g_{\alpha\beta}=\eta_{\alpha\beta}+h_{\alpha\beta}\;,
\end{equation}
where $h_{\alpha\beta}$ is the post-Minkowskian perturbation of the
Minkowski metric tensor $\eta_{\alpha\beta}$. We impose the harmonic
gauge condition \cite{mtw} on the metric tensor
\begin{equation}
\label{hgc}
\partial_\alpha h^{\alpha\beta}-\frac{1}{2}\partial^\beta h^{\lambda}_{\;\,\lambda}=0\;.
\end{equation}
The linearized Einstein equations in the first post-Minkowski
approximation are of the hyperbolic-type (wave) equations for the
metric perturbation. In arbitrary harmonic coordinates $x^\alpha=(ct,
{\bm x})$, the Einstein equations read \footnote{$t$ is
time and ${\bm x}$ is space coordinates.}
\begin{eqnarray}
\label{gfe}
\Box h^{\mu\nu}&=&-\frac{16\pi
G}{c^4}\left(T^{\mu\nu}-\frac{1}{2}\eta^{\mu\nu}T^\lambda_{\;\,\lambda}\right)\;.
\end{eqnarray}
Here $(\Box=-c^{-2}\partial^2/\partial t^2 +\nabla^2)$ is the wave
operator in flat space-time, and $T^{\mu\nu}$ is the stress-energy
tensor of point-like light-ray deflecting bodies in linearized approximation
\begin{eqnarray}
\label{hh} 
T^{\mu\nu}(t, {\bm x})=\sum_{a=1}^N M_a \gamma_a^{-1}(t)\,u_a^\mu(t)\,
u_a^\nu(t)\, \delta^{(3)}\bigl({\bm x}-{\bm x}_a(t)\bigr)\;,
\end{eqnarray}
where the index $a=1,2,...,N$ enumerates gravitating bodies of the
solar system, $M_a$ is the (constant) rest mass of the $a$th body,
${\bm x}_a(t)$ is time-dependent spatial coordinate of the $a$th body,
${\bm v}_a(t)= d{\bm x}_a(t)/dt$ is velocity of the $a$th body,
$u_a^\alpha=\gamma_a (1,\,{\bm v}_a/c)$ is the four-velocity of the
$a$th body, $\gamma_a=\bigl(1-v_a^2/c^2\bigr)^{-1/2}$ is the
Lorentz-factor, and $\delta^{(3)}({\bm x})$ is the 3-dimensional
Dirac's delta-function.  We have neglected $\sqrt{-g}$ in Eq.
(\ref{hh}) because in the linearized approximation $\sqrt{-g}=1+O(h)$
and the terms proportional to the metric tensor perturbation are
irrelevant in $T^{\mu\nu}$.

Because the Einstein equations (\ref{gfe}) are linear, we can consider
their solution as a linear superposition of the solutions for each
body. It allows us to focus on the relativistic effects caused by one
body (Jupiter) only, although in the actual experiment the
gravitational fields of Sun and Earth are also important and were
included in the analysis (see \cite{fk-apj, fk2002} for more detail).
Solving Einstein's equations (\ref{gfe}) by making use of the retarded Li\'enard-Wiechert tensor potentials \cite{bd}, one obtains the post-Minkowski metric
tensor perturbation for Jupiter \cite{bd,ks}
\begin{equation}\label{1} 
h^{\mu\nu}(t,{\bm x})=\frac{2GM_J}{c^2}\frac{2u^\mu u^\nu+\eta^{\mu\nu}}{r_R}\;,
\end{equation}
where $r_R\equiv -u_\alpha r^\alpha$,
$r^\alpha=x^\alpha-x^\alpha_J(s)$.  The metric perturbations for Sun
and Earth look similar to Eq. (\ref{1}) with a corresponding
replacement of masses, coordinates and velocities for the Sun and
Earth respectively.  In Eq. (\ref{1}), $M_J$ is the mass of
Jupiter, $u^\alpha=c^{-1}dx^\alpha_J(s)/d\tau=\gamma(s)(1, {\bm
v}_J(s)/c)$ is its four-velocity, with $\tau=\tau(s)$ being a proper
time on Jupiter's world line, ${\bm v}_J(s)=d{\bm x}(s)/ds$ is
Jupiter's coordinate velocity, and
$\gamma(s)=ds/d\tau=(1-v_J^2(s)/c^2)^{-1/2}$. Notice that the metric
tensor perturbation (\ref{1}) is valid for accelerated motion of
Jupiter and is not restricted by the approximation of a body moving on
a straight line (see \cite{bd} for more detail). In other words, the
four-velocity $u^\alpha$ in Eq. (\ref{1}) is not a constant.

Because we solved the Einstein equations
(\ref{gfe}) in terms of the retarded Li\'enard-Wiechert potentials, the distance $r^\alpha=x^\alpha-x^\alpha_J(s)$, the
Jupiter's worldline $x^\alpha_J(s)=(cs, {\bm x}_J(s))$, and the
four-velocity $u^\alpha(s)$ are all functions of the {\it retarded} time
$s$ \cite{bd}. The retarded time $s$ is found in the first post-Minkowski approximation as a solution of the {\it gravity null cone} equation
\begin{equation}
\label{grav}
\eta_{\mu\nu}r^\mu
r^\nu\equiv\eta_{\mu\nu}\Bigl(x^\mu-x^\mu_J(s)\Bigr)\Bigl(x^\nu-x^\nu_J(s)\Bigr)=0\;,
\end{equation}
that is
\begin{equation}
\label{1a}
s=t-\frac{1}{c}|{\bm x}-{\bm x}_J(s)|\;,
\end{equation}
where the fundamental constant $c$ in Eq. (\ref{1a}) must be
conceptually understood as the speed of characteristics of the gravity
field equations because it comes from the wave operator of the
Einstein gravity field Eqs. (\ref{gfe}).  Indeed, Maxwell
equations have not yet been considered; hence the retardation in
Eq. (\ref{1a}) cannot be associated with the propagation of
electromagnetic waves (light). In what follows, the retarded time $s$
will appear in all subsequent equations related to the propagation of
radio waves. It is important to keep in mind that Eq. (\ref{1a})
connects the point of observation ${\bm x}$ and the retarded position
of Jupiter ${\bm x}(s)$ by a null line which is always a
characteristic of the Einstein field equations. Radio waves are also
propagating along a null direction connecting the observer and the
source of radio waves (quasar). However, the null characteristic of
the Einstein equations (\ref{1a}) is well separated in the sky from the
null direction associated with the propagation of the radio wave so
that they can not be observationally confused (see Appendix A for additional details).

\subsection {Interpretation of the Einstein equations}

The retardation shown in Eqs. (\ref{1}) -- (\ref{1a}) is
physically associated with the finite speed of propagation of the null
characteristics of the gravity field. In the most general case, this
propagation is realized in the form of a traveling gravitational field
which can be decomposed into components of different algebraic types
\cite{pir}. In the wave zone of the solar system, the traveling
gravitational field decouples from the system and propagates as a
transverse-traceless (TT) gravitational wave with amplitude decaying
as $1/r$ where $r$ is the distance from the system. However, in the
near and transition zones of the solar system, a propagating
gravitational field has components which are not completely decoupled
from the matter and decay faster than $1/r$ \cite{pir}. These
components of the propagating gravitational field are closely
associated with the residual gauge freedom of the metric tensor
$g_{\alpha\beta}$ and for this reason we call them the gauge
modes. They are also known as transverse (T) and longitudinal (L) wave
modes \cite{mtw}. These gauge modes control the validity of the laws
of conservation of matter and gravity field in the lower order
post-Newtonian approximations \cite{carl,ksge}.
  
   All components of the propagating gravitational field of the solar
system interact with radio waves moving from the quasar to the Earth,
and perturb the phase of an electromagnetic waves according to the
retardation Eq. (\ref{1a}) which incorporates both the near-field
and far-field effects of gravity as explained in \S 3.  The use of the
retarded Li\'enard-Wiechert potentials, rather than the advanced
potentials, is consistent with the principle of causality, and the
observation of the orbital decay of binary pulsars caused by emission of gravitational
radiation \cite{wt}.  

The speed $c$ is a fundamental speed of the special theory of
relativity and Lorentz-invariance of flat (Minkowski) space-time
metric $\eta_{\alpha\beta}$. According to
our definition the {\it speed of gravity} is a fundamental constant
of the curved space-time manifold which is associated with
temporal variation of gravitational field and appears as a physical quantity any time when one takes a
derivative of the metric tensor with respect to time coordinate $x^0$
\cite{kop-cqg}.  In GR the speed of gravity is equal to $c$ because at each point the curved
space-time manifold has a tangent space-time which coincides with Minkowski space. But this postulate must be tested experimentally
\cite{kop-cqg,w-book}.

   The measured Jovian deflection is most sensitive to the retardation
of the gravitational field caused by its Lorentz invariant properties.  Thus, what we measured was not the speed of TT
gravitational waves but the speed of 
gravity in the near-field zone.  Since the speed of TT gravitational waves can
not exceed the speed of gravity in the near-zone
for any viable theory of gravity \cite{bum}, our VLBI experiment sets an upper limit on
the speed of propagation of gravitational waves as well. 

\subsection{The gravito-magnetic field}

Einstein's equations in the frame attached to Jupiter do not contain
time derivatives of the metric tensor because the Jovian field is
static; hence the retardation of gravity (associated with time
derivatives of the metric tensor and its null characteristics) cannot
be measured in this frame.  This condition has led many to infer that
the null characteristics and retardation of the gravity field must
remain unmeasurable after a Lorentz transformation to another frame,
for example to the surface of the earth or the barycenter of the solar
system.  Thus, it is claimed, that the deflection-of-light measurement
cannot be associated with the retardation of gravity.  But, the
Lorentz transformation of the Einstein equations (\ref{gfe}) from static to a
moving frame induces time derivatives of the metric tensor perturbations which can be measured in the light deflection
experiments from the retardation of gravity field \cite{fk-apj} in
order to determine the fundamental speed of its null characteristics.

Our interpretation does not mean that the laws of gravitational physics change under
a Lorentz transformation, but only that some property of gravity can
become measurable after transformation from a static to a non-static
frame.  A similar example comes from electrodynamics in the case of a
single electron which has a static electric field in its own frame.  A
Lorentz transformation to a moving frame induces a new field -- the
magnetic field, which was not observable in the static frame.  
Maxwell's equations, also, show that the strength of the magnetic
field in the non-static frame depends on a time derivative of the
electric field coupled with the speed of light.

The same analogy is associated with the Jovian deflection experiment.
In Jupiter's frame there is only a static ``electric-type"
gravitational field.  With a Lorentz transformation to a moving frame,
a ``magnetic-type" field, called a {\it gravito-magnetic} field, is
induced.  This field, according to the theory of general relativity,
arises from moving matter (matter currents) just as an ordinary
magnetic field arises from moving charges (electrical currents). The
analogy is so apt that the equations describing this ``magnetic-type"
component of gravity can essentially be adapted from Maxwell's
equations for electromagnetism by replacing the charge density with
the mass density and the charge current with the mass current, with
changes of numerical factors reflecting different helicities of
electromagnetic (spin 1) and gravitational (spin 2) fields
\cite{mash}.  Thus, in a moving frame, the translational motion of
Jupiter produces the gravito-magnetic field which deflects the light tangentially
by dragging it to the direction of motion of Jupiter \cite{kop-cqg,
kop-pla}.  We can measure this tangential gravito-magnetic dragging of light and
express its magnitude in terms of the speed of characteristics of the
gravity field equations, in the analogous
way that we can express the magnitude of the magnetic field
interaction in terms of the speed of the characteristics of
electromagnetic field equations
\cite{kop-cqg, kop-pla}.

However, the predicted light deflection must be invariant to any
Lorentz transformation.  This is clearly the case because all of the
calculations of the deflection experiment, regardless of the reference
frame, predict the same deflection!  But, the gravito-magnetic
property of gravity that is sampled by the experiment does depend on
the reference frame.  In Jupiter's frame, all of the deflection is
associated with the static gravitational field -- but we do not
observe in this frame.  In the frame of the experiment (the
barycentric of the solar system), the deflection is associated with a
(radial) static gravitational field of Jupiter plus a small (tangential)
gravito-magnetic field, due to its orbital velocity, which is coupled
with the speed of gravity, and reveals itself through the retardation
of gravity already in terms of order $v/c$ beyond the Shapiro delay.
More precise measurements of the deflection by Jupiter, which
are directly sensitive to TT gravitational waves, will require a
more exact solution of the Einstein equations taking into account
Jupiter's orbital acceleration.

\subsection{The speed of gravity and the Newtonian limit}

The Principle of Correspondence of GR with the Newtonian theory (closely
related to the Principle of Equivalence) does not provide any information
on the speed of gravity because the speed of gravity is
associated with the temporal changes of the gravitational field while
the Newtonian limit is stationary.  In principle, the speed of
gravity should appear in the right side of the Einstein equations as a
constant characterizing coupling between space-time and matter
variables \cite{kop-cqg}. Thus, it seems natural to state that the
speed of gravity can be already derived from the validity of the
Principle of Correspondence between Einstein's theory of general
relativity and the Newtonian theory of gravity.  But, the metric
tensor perturbation in the Newtonian limit of general relativity
depends on a coupling constant $\kappa=G/c^2$, where $G$ is the
observed value of the universal gravitational constant.  If the speed
of gravity were different from $c$ the difference would be absorbed by
re-definition of the {\it
observed} value of $G$ in such a way that the coupling constant
$\kappa$ is kept fixed \cite{kop-cqg, w-book}.

\section{The Deflection Experiment}

\subsection{The electromagnetic phase}

Let us assume that the speed of light is equal to the speed of gravity. It simplifies calculations but the observable effects of the light and gravity propagation can be still clearly separated as we discuss later. For the doubting reader appendix of this paper gives more general covariant description of propagation of light rays in a bi-metric theory of gravity, thus, helping one to formally discern null characteristics of light from those for gravity given by Eq. (\ref{1a}).

Very Long Baseline Interferometry (VLBI) measures the phase $\varphi$
of the wave front coming from a radio source.  The phase is an
invariant scalar function with respect to coordinate transformations,
and it is determined in the approximation of geometric optics from the
eikonal equation \cite{mtw,ll}
\begin{equation}
\label{eik}
g^{\mu\nu}\partial_\mu\varphi\partial_\nu\varphi=0\;,
\end{equation}
where $g^{\mu\nu}=\eta^{\mu\nu}-h^{\mu\nu}$. 
The eikonal Eq. (\ref{eik}) is a
direct consequence of Maxwell's equations \cite{mtw,frol,km} and its
solution describes a wave front of an electromagnetic wave propagating
in curved space-time defined by the metric tensor in Eqs.
(\ref{y}), (\ref{1}). We emphasize that the electromagnetic wave in
Eq. (\ref{eik}) is a test field that has no back-action on the
properties of the metric tensor perturbation $h_{\mu\nu}$, and does
not change the curvature of the space-time. Hence, the structure of Eq.
(\ref{eik}) is not related to the specific properties of the
electromagnetic field, but is completely determined by the structure
of space-time, or, more accurately, its geometry defined by Eqs.
(\ref{y}), (\ref{1}) \cite{frol}.

Let us introduce a co-vector $K_\alpha=\partial_\alpha\varphi$. Let
$\lambda$ be an affine parameter along a light ray orthogonal to the
electromagnetic wave front $\varphi$. The vector
$K^\alpha=dx^\alpha/d\lambda=g^{\alpha\beta}\partial_\beta\varphi$ is
tangent to the light ray. Eq. (\ref{eik}) expresses a simple fact
that the vector $K^\alpha$ is null, that is $g_{\mu\nu}K^\mu
K^\nu=0$. Thus, the light rays are null geodesics defined in the
linearized post-Minkowskian approximation by equation \cite{ll}
\begin{equation}
\label{geo}
\frac{dK_\alpha}{d\lambda}=\frac{1}{2}\partial_\alpha h_{\mu\nu}K^\mu K^\nu\;,
\end{equation}  
which describes propagation of a light (radio waves) ray from the
quasar to the observer (VLBI station).  Wave Eq. (\ref{eik}) and ray Eq.
(\ref{geo}) are supplementary to each other and have equivalent
physical content.

A more straight-forward solution can be
obtained to Eq. (\ref{eik}).  If Jupiter's acceleration is neglected, a plane-wave solution of this equation is
\begin{equation}
\label{3}
\varphi(x^\alpha)=\varphi_0+k_\alpha x^\alpha+\frac{2GM_J}{c^2}\left(k_\alpha 
u^\alpha\right)\ln\left(-k_\alpha r^\alpha\right)\;,
\end{equation}
where $\varphi_0$ is a constant of integration,
$k^\alpha=c^{-1}(\nu,\nu{\bm k})$ is a constant wave vector of the
unperturbed radio wave such that $\eta_{\mu\nu}k^\mu k^\nu=0$, $\nu$ is its unperturbed constant frequency.  The speed $c$ in the
expression for the wave vector $k^\alpha$ must be conceptually understood as the
speed of light, and the unit vector ${\bm k}$ defines the unperturbed
direction of propagation of the radio wave from the quasar.
One can easily prove that Eq. (\ref{3}) is a particular solution
of the electromagnetic eikonal Eq. (\ref{eik}). Indeed, observing
that
\begin{equation}
\label{qa}
\partial_\alpha r^\mu=\delta^\mu_\alpha-\frac{u^\mu}{\gamma}\partial_\alpha s\;, 
\end{equation}
where $\gamma=(1-v_J^2/c^2)^{-1/2}$, one obtains from the gravity null cone Eq. (\ref{grav})
\begin{equation}
\label{po} 
\partial_\alpha s=-\gamma\,\frac{r_\alpha}{r_R}\;.
\end{equation}
Differentiation of Eq. (\ref{3}) using Eqs. (\ref{qa}) and
(\ref{po}) shows that the eikonal Eq. (\ref{eik}) is satisfied.

Eq. (\ref{3}) for the electromagnetic phase is clearly
Lorentz-invariant and valid in an arbitrary coordinate system.  The
gravity field (transverse-traceless and gauge modes) propagates from
Jupiter along the hypersurface of gravity null cone (\ref{1a}) and
perturbs the phase front of the radio wave at a field point $x^\alpha$
(see figure \ref{null}).  Notice that the procedure of solving of the
eikonal Eq. (\ref{eik}) does not change the physical meaning of
the retarded time $s$ in the coordinate of Jupiter $x^\alpha_J(s)$,
affected by the propagation of the gravity. Eq. (\ref{1a}) {\it
does not} depend on the wave vector $k^\alpha$ which alone contains
the information about the speed and direction of propagation of the
quasar's radio wave. In other words, a gravitating body (Jupiter)
always perturbs the electromagnetic phase and deflects the light by
acting from its retarded position ${\bm x}_J(s)$ regardless of the
direction of motion of the incoming photon and/or the magnitude of its
impact parameter with respect to the light-ray deflecting body. Hence, the speed of the physical light (radio waves) used for VLBI observations can not enter in any way to the gravity null cone Eq. (\ref{1a}). The light is used as a test particle in order to measure the retardation of gravity effect through the retarded position of Jupiter ${\bm x}_J(s)$ which is determined in the experiment from the magnitude and direction of the gravitational deflection of light and is compared with the JPL ephemeris position of Jupiter obtained independently from its direct observation in a series of radio-tracking and/or VLBI measurements of spacecrafts which orbited Jupiter \cite{standish} (see Fig. \ref{twocones} for comparison of the two positions of Jupiter).  

    This independence of the retarded time $s$ from the
electromagnetic wave vector $k^\alpha$ clearly shows that the
retardation is not due to the propagation of the radio wave from the
quasar. The radio propagation can not be responsible for the
retardation effect in Jupiter's position because the radio wave propagates in a different direction (see Fig. \ref{vlbi}). The retardation in Jupiter's position is a consequence of
the finite value of the fundamental speed of propagation of gravity and exists independently of solving
the eikonal equation; hence the solution of the eikonal equation can
not change this property of the propagating gravity field. This
conclusion is supported in more detail in \cite{kop-cqg,kop-ijmp1,kop-ijmp2,kop-pla2}.

\subsection{The gravitational time delay}

The Lorentz-invariant relativistic time delay equation, generalizing the
static Shapiro delay \cite{shap}, can be obtained directly from
Eq. (\ref{3}). We note that the phase $\varphi$ of electromagnetic wave,
emitted at the point $x^\alpha_0=(ct_0,{\bm x}_0)$ and received at the
point $x^\alpha=(ct,{\bm x})$, remains constant along the wave's path
\cite{mtw,ll,km}. Indeed, if $\lambda$ is an affine parameter along
the path, one has for the phase's derivative
\begin{equation}
\label{phase}
\frac{d\varphi}{d\lambda}=\frac{\partial\varphi}{\partial x^\alpha}\frac{
dx^\alpha}{d\lambda}=K_\alpha K^\alpha=0\;,
\end{equation}
which means that $\varphi\left(x^\alpha(\lambda)\right)= ${\rm const} in accordance with our assertion.  Equating two values of the phase at the points
$x^\alpha_0$ and $x^\alpha$ and separating time and space coordinates
one obtains from (\ref{3})
\begin{equation}
\label{tde}
t-t_0=\frac{1}{c}{\bm k}\cdot\left({\bm x}-{\bm
x}_0\right)-\frac{2GM_J}{c^3}\frac{1-c^{-1}{\bm k}\cdot{\bm
v}_J}{\sqrt{1-v^2_J/c^2}}\ln\left(r-{\bm k}\cdot{\bm
r}\right)+\mbox{\rm const}\;,
\end{equation}
where `{\rm const}' denotes all of the constant terms and the
relativistic terms taken at the emitting point, the distance ${\bm
r}={\bm x}-{\bm x}_J(s)$, $r=|{\bm x}-{\bm x}_J(s)|$, and the retarded
time $s$ is defined by the gravity cone Eq. (\ref{1a}). The  Lorentz-invariant expression for time
delay (\ref{tde}) was derived first by Kopeikin and Sch\"afer
\cite{ks} who solved equations for light geodesics in the
gravitational field of moving bodies. Klioner \cite{kl} also obtained
this expression by making use of the Lorentz transformation of Jupiter's coordinate and the light vector in the
Shapiro delay (which is implicitly equivalent to transforming the gravity and electromagnetic field equations) from a
static to moving frame. Notice that in GR Eq. (\ref{tde}) describes a
hypersurface of both the light and gravity null cones along which electromagnetic field from
the quasar and the gravitational field from Jupiter are propagating respectively (see Fig. \ref{null}). The light cone intersects with the
gravity null cone given by Eq. (\ref{1a}) at the point $x^\alpha$ at which the
propagating Jupiter's gravity field reaches the hypersurface of
electromagnetic phase at time $t$ when Jupiter is located at its
retarded position ${\bm x}_J(s)$ coinciding with the origin of the
retarded gravity null cone.  Additional details are shown in appendix where we used a bi-metric theory of gravity. In this theory the light null-cone
does not coincide with the gravity null-cone as it was in GR, and it implies that light can not propagate on the
hypersurface of the gravity null cone (\ref{1a}). Hence, the retarded position of Jupiter measured in
Jovian experiment is due to the finite value of the fundamental speed of gravity and it sets the upper limit on the speed of
propagation of gravitational waves which can propagate either on shell
of the gravity null-cone (\ref{1a}) in case of GR \cite{bum} or inside it in the
case of the theory of gravity with massive gravitons \cite{grish}.

\subsection{The aberration of gravity}

Let us start with the light geodesics Eq. (\ref{geo}). Taking
partial derivative from the metric tensor recasts Eq. (\ref{geo})
to the following form
\begin{equation}
\label{j}
\frac{dK_\alpha}{d\lambda}=\frac{2GM_J}{c^2}(k_\mu u^\mu)^2\partial_\alpha\left(\frac{1}{r_R}\right)=-\frac{2GM_J}{c^2}\frac{(k_\mu u^\mu)^2}{r_R^2}\, n_\alpha\;,
\end{equation}  
where \cite{bd}
\begin{equation}
\label{x}
n^\alpha\equiv\partial^\alpha r_R = \frac{r^\alpha}{r_R}- u^\alpha\;,
\end{equation}
and we have neglected Jupiter's acceleration. 

The vector $r^\alpha$ is null due to the gravity null cone Eq.
(\ref{grav}), but the vector $n^\alpha$ is space-like, because
$n_\alpha n^\alpha=+1$. This occurs because the retardation of gravity
effect in the first term in the right side of Eq. (\ref{x}) (present in the coordinate ${\bm x}_J(s)$
of Jupiter through the retarded time $s$) is compensated by the second
term depending on the four-velocity of Jupiter $u^\alpha$ which is a
time-like vector, $u_\alpha u^\alpha=-1$.  This velocity-dependent
term present in the gravitational force is associated in GR with the
relativistic effect known as aberration of gravity \cite{carl} because
it describes an aberration-like change in the direction of the
gravitational force (that is, from the null-like vector $r^\alpha/r_R$ to
the space-like vector $n^\alpha$) caused by motion of the gravitating
body with respect to the reference frame used for data reduction.

The compensation of the retardation of gravity by the aberration of
gravity in the expression for vector $n^\alpha$ can be demonstrated
using a Taylor (post-Newtonian) expansion of $x^\alpha_J(s)$ around
time $t$.  This gives
\begin{eqnarray}
\label{oo}
r^\alpha&=&R^\alpha-(s-t)\frac{dx^\alpha}{ds}=R^\alpha+\frac{r}{c}\frac{u^\alpha}{\gamma}\;,
\\\label{vv}
r_R&=&-u_\mu R^\mu+\frac{r}{c\gamma}\;,
\end{eqnarray}  
where terms depending on Jupiter's acceleration have been neglected,
$R^\alpha=x^\alpha-x^\alpha_J(t)=\bigl(0,{\bm x}-{\bm x}_J(t)\bigr)$ is a purely
spatial vector lying on the hypersurface of constant time $t$,
$r=|{\bm x}-{\bm x}_J(s)|$, and the retarded time Eq. (\ref{1a})
has replaced $s-t$ with $r/c$ where $c$ in this case is understood as the
fundamental speed of gravity. Substitution of Eqs. (\ref{oo}), (\ref{vv})
into Eq. (\ref{x}), and reduction of similar terms, show that
retarded term proportional to $s-t$ cancels with the gravity
aberration term proportional to $u^\alpha$ and Eqs. (\ref{x}) and (\ref{vv}) are reduced to
\begin{eqnarray}
\label{j1}
n^\alpha&=&\left[R^\alpha+u^\alpha \left(u_\mu R^\mu\right)\right]r_R^{-1}\;,
\\\label{poq}
r_R&=&\left[R_\alpha R^\alpha+(u_\alpha R^\alpha)^2\right]^{1/2}\;.
\end{eqnarray}

This result, shown elsewhere \cite{carl}, reveals that (in the first approximation with respect to the four-velocity $u^\alpha$) the
gravitational acceleration of a light particle (photon) is directed
toward the {\it present} position of Jupiter ${\bm x}_J(t)$ separated
from the photon by a space-like vector $R^\alpha$.  At first glance,
this might imply that the aberration of 
gravity cannot be measured in the linear $v_J/c$ approximation beyond
the Shapiro delay. This point of view is advocated in \cite{ass,will-apj,carlip,samuel,pask}.  

However, it is crucial to understand that the VLBI experiment does not measure the gravitational acceleration (the right side of Eq. (\ref{j})) of radio
wave (light particles) which is not VLBI-measurable quantity. What VLBI measures in reality is the
electromagnetic phase of the radio wave from a quasar and direction of its
propagation, which do depend on the retarded position of Jupiter
${\bm x}_J(s)$ in contrast to the gravitational acceleration. In order to see how it happens let us integrate the inverse distance $1/r_R$ along the light-ray trajectory parametrized by the affine parameter $\lambda$. It results in
\begin{equation}
\label{br}
\frac{d\lambda}{r_R}=-\frac{1}{\gamma}\,\frac{ds}{k_\alpha r^\alpha}=\frac{1}{k_\alpha u^\alpha}\,d\Bigl[\ln\left(-k_\alpha r^\alpha\right)\Bigr]\;,
\end{equation}
which can be easily confirmed by direct differentiation of the gravity null cone
Eq. (\ref{grav}) taken on the trajectory of the radio wave signal
$x^\alpha=x^\alpha(\lambda)$.  
Using Eq. (\ref{br}) for the
integration of the light geodesic Eq. (\ref{j}) gives
\begin{equation}
\label{zb}
K_\alpha=k_\alpha+\frac{2GM_J}{c^2}\frac{k_\mu u^\mu}{k_\beta r^\beta}
\left[k_\alpha+(k_\nu u^\nu)\frac{r_\alpha}{r_R}\right]\;,
\end{equation}
which coincides with the expression for
$K_\alpha=\partial_\alpha\varphi$, derived by differentiation of the
electromagnetic phase (\ref{3}).  The subsequent post-Newtonian expansion (\ref{oo})
of $r^\alpha$ in the function $k_\alpha r^\alpha$, that enters solution
(\ref{zb}) for $K_\alpha$ and that (\ref{3}) for the phase
$\varphi(x^\alpha)$, gives terms which depend on the aberration and
the fundamental speed of gravity already in linear order $\sim v_J/c$. This result is
confirmed in \cite{kop-cqg,kop-ijmp1} using general relativity parametrized by a single speed-of-gravity parameter $c_g$ characterizing all non-stationary effects of gravitational field.

\subsection{Why can VLBI determine the retarded position of Jupiter?\label{rp}}

The gravitational force exerted by moving Jupiter on any mass
(including radio photons) is a spatial vector directed in the first approximation toward the present position of
Jupiter as shown in Eqs. (\ref{j}) and (\ref{j1}).  A question
arises: ``How can the gravitational force be a spatial vector but the phase of electromagnetic waves, measured
by VLBI, be gravitationally affected from null direction given by Jupiter's retarded position?" The answer to this
question arises from the nature of gravity and radio waves which both move along null
rays and their wave fronts form a null hypersurface (null cone).

The gravitational 4-force on the right side of Eq. (\ref{j}) has
a Newtonian analog, and for this reason this force is a space-like
vector directed to the present position of Jupiter $x^\alpha_J(t)$.
Integration of the 4-force along particle's trajectory gives a vector of a
linear momentum of the particle which is tangent to particle's world
line. In the case of a slowly moving particle, its linear momentum is a time-like vector. However, radio photons move along
light geodesics and their linear momentum is a null wave vector
\cite{mtw,ll}. Hence, the integration of the light geodesics (\ref{j})
along the photon trajectory must produce the null vector
$K^\alpha=k^\alpha+\delta k^\alpha$ of the photon, where $k^\alpha$ is
an unperturbed wave vector of the photon and $\delta k^\alpha$, given by Eq. (\ref{zb}), is its
gravitational perturbation caused by the space-like vector of the gravitational
force in the right side of the equations of motion (\ref{j}).

The null vector $K^\alpha$ is a gradient of the radio phase with
$K^\alpha=g^{\alpha\beta}\partial_\beta\varphi$ and the phase
$\varphi(x^\alpha)$ is a null hypersurface. The 
gravitational field of Jupiter also propagate along null characteristics 
defined by vector $r^\alpha=x^\alpha-x^\alpha_J(s)$, where
$x^\alpha_J(s)$ is a retarded position of Jupiter taken at the
retarded time $s$ defined by the gravity null cone Eq.
(\ref{1a}). For this reason, the characteristics of the
gravitational field of moving Jupiter form a null hypersurface,
gravitational phase $\varphi_{gr}(x^\alpha)$, as well. Radio photon
interacts with the gravitational force at point $x^\alpha$. Therefore,
two null hypersurfaces $\varphi(x^\alpha)$ and
$\varphi_{gr}(x^\alpha)$ intersect at this point and form a null
sub-hypersurface such that both vectors $K^\alpha$ and $r^\alpha$
belong to it, and the gravitational perturbation of the radio phase,
$\varphi(x^\alpha)$, is defined by the structure of this null
sub-hypersurface.

There are only three scalars that can be formed from the three vectors
$k^\alpha$, $r^\alpha$, and $u^\alpha$ (Jupiter's velocity): $k_\alpha
u^\alpha$, $r_R=-u_\alpha r^\alpha$, and $k_\alpha r^\alpha$, and the
(weak) perturbation of the radio wave phase can depend on a linear
combination of some functions having these three scalars as
arguments.  The scalar $r_R$, however, can be eliminated because the
partial derivative of the phase $\partial_\alpha\varphi$ must be a
null vector, whereas the partial derivative $\partial^\alpha r_R
=n^\alpha$ (see Eq. (\ref{x})) is a space-like vector which does
not lie on the null hypersurface. The scalar $k_\alpha u^\alpha$ is
clearly a constant, if one neglects Jupiter's acceleration, so that $\partial^\beta(k_\alpha u^\alpha)=0$. Thus, term $k_\alpha u^\alpha$ has no influence on the propagation direction
of the photons.  We conclude that the only significant scalar combination
defining the gravitational perturbation of the electromagnetic phase
$\varphi$ is simply $k_\alpha r^\alpha$, which depends on the retarded
position $x^\alpha_J(s)$ of Jupiter taken at the retarded instant of
time $s$ defined by the {\it gravity null cone} Eq. (\ref{1a}) obtained as a retarded solution of the gravity field Eqs. (\ref{gfe}).
The exact solution of the eikonal Eq. (\ref{eik}) confirms the
above argument, and shows explicitly that the radio phase $\varphi$ is sensitive to the
retarded position of Jupiter ${\bm x}_J(s)$, whereas the gravitational force, acting
on photons as a space-like vector, is sensitive to the present
position of Jupiter ${\bm x}_J(t)$.

In summary, the solution of the Einstein equations shows that gravity force
propagates from a moving massive body; hence the retarded position of
the body, if the speed of gravity propagation is not infinite, is a
natural consequence of the theory.  But, the interaction of gravity in
the dynamics of slowly moving bodies ($v \ll c$) will not produce a
null vector since their four-velocities are not partial derivatives of a null hypersurface, and the retardation of gravity effect
is relegated to higher orders of $v/c$ in the solutions of the
equations of motion for these bodies \cite{kop85,gr-kop,damour}. On the other hand,
the phase (or time delay) of the radio waves which is measured by
VLBI, is sensitive to the retardation of gravity effect (retarded
position of Jupiter) at order $v/c$ because in GR gravity has the same speed as light which means that the fundamental speed in the Maxwell and Einstein equations is one and the same. This prediction of general relativity was the goal of experimental testing in Jovian experiment.

\subsection{Fundamental speed of gravity as a fitting parameter}\label{sogas}

Our theoretical formulation of the problem of propagation of light in
a time-dependent gravitational field of moving bodies predicts that
the retarded coordinate of Jupiter ${\bm x}_J(s)$ in all terms depending on gravity field in Eqs.
(\ref{1})--(\ref{3}) is shifted from its present position ${\bm
x}_J(t)$ due to the finite time the gravity takes to propagate from a
moving massive body (Jupiter) to the field point $x^\alpha$. To quantify this
prediction experimentally we have developed a co-ordinated
fiber-bundle parameterization of both the Einstein equations and light
geodesics, based on a (single) speed-of-gravity parameter $c_g$. This $c_g$-parametrized approach is explained in our papers
\cite{kop-cqg,kop-ijmp1} \footnote{Appendix of the present paper gives another theoretical parametrization of the speed of gravity versus the speed of light}. This $c_g$-parameterization decouples the null characteristics of electromagnetic and gravitational field already at the first order terms in $v/c$ in contrast to
the PPN formalism by Nordtvedt and Will \cite{will-apj} which starts to distinguish the light and gravity null cones only in terms of the second order in $v/c$ \cite{kop-ijmp1}.  We assert that
on each bundle of the parametric general-relativistic space-time we are allowed to substitute
Eq. (\ref{1a}) with the following description of the
characteristics of the gravity field equations
\begin{equation}
\label{2}
s_g=t-\frac{1}{c_g}|{\bm x}-{\bm x}_J(s_g)|\;,
\end{equation}
where $c_g$ is a {\it speed of gravity} parameter running formally
from $c_g=\infty$ to $c_g=c$.  Again, we assert that in the papers \cite{kop-cqg,kop-ijmp1} we are not using
$c_g\not=c$ in the sense of an alternative theory of gravity. We work
with various values of $c_g$ in the fiber-bundle space-times of GR
from which only one coincides with physical space-time when $c_g=c$.
Parameter $c_g$ must be considered rather as a label marking matrix of the Lorentz transformation of the gravitational field variables \cite{kop-ijmp1}, making it clearly different from the matrix of the Lorentz transformation of electromagnetic field which depends on the speed of light $c$. For this reason $c_g$ is not a measure of
the PPN parameter $\alpha_1$ which is associated with preferred frame-velocity
tests of GR, described in \cite{w-book} (see misconception 7).

Values of $c_g$ less than $c$ (although such values can be formally
obtained from the fitting procedure) are  strongly restricted by
observations of cosmic rays. Indeed, if $c_g$ were significantly
less than $c$, the cosmic rays would radiate energy mostly in
the form of gravitational Cherenkov's radiation, which contradicts the
observations \cite{mn}.  Eq. (\ref{2}) is valid up to the
post-Newtonian terms of second order in the sense that it is obtained
as a retarded solution of the Einstein equations where all time
derivatives contain a single speed-of-gravity parameter $c_g$
\cite{kop-cqg,kop-ijmp1}. The value of the parameter $c_g$ in our model of
the experiment was found by minimizing the phase residuals
$\delta\varphi=|\varphi_{\rm obs}-\varphi_{\rm cal}(x^\alpha)|$, where
$\varphi_{\rm obs}$ is the observed value of the phase, and
\begin{equation}
\label{3a}
\varphi_{\rm cal}(x^\alpha)=\varphi_0+k_\alpha x^\alpha+\frac{2GM_J}{c^2}\left(k_\alpha 
u^\alpha\right)\ln\left(-k_\alpha \rho^\alpha\right)\;,
\end{equation}
which is the model function for the phase obtained from Eq.
(\ref{3}) after mathematically legitimate replacement of $r^\alpha$ with
$\rho^\alpha=x^\alpha-x^\alpha_J(s_g)$ \cite{kop-cqg,kop-ijmp1}.

We have introduced the speed-of-gravity parameter $c_g$ to Eqs.
(\ref{2}) and (\ref{3a}) in such a way that eikonal Eq.
(\ref{eik}) does not change the physical identity of the gravity null
cone Eq. (\ref{1a}) which treats the retardation as due to the
finite speed of the gravity field.  The
solution to the Einstein equations remains the same as in Eq.
(\ref{1}) with $s$ replaced with $s_g$ \cite{kop-cqg,kop-ijmp1}.  This $c_g$ {\it
speed-of-gravity} parameterization of the Einstein equations demands
that the speed of the gravitational field is the same in both the wave
and near zones of an isolated system emitting gravitational waves.  Our experimental goal was to measure $c_g$ in the near zone, and to determine if
$c_g=c$ predicted by general relativity. As we noted previously, the
experiment can not measure the speed of gravitational waves in the wave zone directly
but the measurement of $c_g$ in the near zone definitely imposes the upper limit on the
speed of their propagation everywhere.

The retardation of gravity is unobservable in the solutions of the
post-Newtonian equations of motion of self-gravitating bodies (a
binary pulsar and the Earth-Sun, for example) in terms of order $v/c$
and $v^3/c^3$ beyond the Newtonian law of gravity because of a perfect
cancellation of the retarded and velocity-dependent terms present in
the gravitational force between the bodies \cite{kop85,gr-kop,damour}.  However, the
present paper (see also \cite{kop2001,kop-ijmp2,kop-pla2}) shows that the
retarded terms are not canceled in the solution (\ref{3}) of the eikonal
equation by the velocity-dependent terms. The phase
$\varphi$ has an argument $k_\alpha r^\alpha$ which does not depend on
the velocity of Jupiter explicitly.  The post-Newtonian
expansion of $k_\alpha r^\alpha$ around the present time $t$
generates velocity-dependent terms which are physically explained by the finite speed of gravity.  

Our $c_g$ parametrization \cite{kop-cqg,kop-ijmp1} assumes that the speed-of-gravity parameter $c_g$ appears in the equation for the time delay $t-t_0$ taken by light to propagate from ${\bm x}_0$ to ${\bm x}$ in the following form
\begin{equation}
\label{dfg}
t-t_0=\frac{c^3}{2\nu^2}\int_{t_0}^t d\zeta\int_{-\infty}^{\zeta}k^\mu k^\nu\left[k^i\,\frac{\partial h_{\mu\nu}(\tau,{\bm x})}{\partial x^i}+\frac{1}{c_g}\frac{\partial h_{\mu\nu}(\tau,{\bm x})}{\partial\tau}\right]_{{\bm x}=c{\bm k}(\tau-t_0)+{\bm x}_0}d\tau\;,
\end{equation}
where $k^\alpha$ is a wave vector of light, $\nu$ is its frequency, and ${\bm k}=(k^i)$ is a unit vector along the direction of its unperturbed propagation. Integration of Eq. (\ref{dfg}) leads exactly to formula (\ref{tde}) with the retarded time $s$ replaced with $s_g$ defined in Eq. (\ref{2}) of this section. It is clear that in case $c_g=\infty$ the Jovian deflection experiment would not measure the integral from the time derivative of the metric tensor shown as the second (gravito-magnetic) term in the right side of Eq. (\ref{dfg}). However, we have measured $c_g\simeq c$ with the precision 20\% and, thus, proved that the experiment is sensitive to the gravito-magnetic dragging of light generated by the partial time derivative of the metric tensor \cite{kop-cqg,kop-ijmp1}. 

\subsection {The determination of the speed of gravity}\label{detcg}

The speed-of-gravity parameter was determined from the experiment in the
following way.  Let us introduce two angles $\Theta=\Theta(t)$ and
$\theta=\theta(s_g)$. The angle $\Theta$ is between the unit vector
${\bm k}$ characterizing direction of propagation of the light ray and
the unit vector ${\bm p}={\bm R}/|R|$ while the angle $\theta$ is
between vector ${\bm k}$ and ${\bm l}={\bm
l}(s_g)={\bm\rho}/\rho$. Here ${\bm R}={\bm x}-{\bm x}_J(t)$ connects
the present position of Jupiter ${\bm x}_J(t)$ at the time of observation
$t$ and the point of observation ${\bm x}$, whereas ${\bm\rho}={\bm
x}-{\bm x}_J(s_g)$ connects the retarded position of Jupiter at the
retarded time $s_g=t-\rho/c_g$ and the point ${\bm x}$ (see Eq.
(\ref{2})).  By definition $\cos\Theta={\bm k}\cdot{\bm p}$, and
$\cos\theta={\bm k}\cdot{\bm l}$ (see figure \ref{vlbi}).

From the definitions of vector ${\bm l}(s_g)$ and the angle
$\theta(s_g)$, we can obtain their post-Newtonian expansions in
the near zone of the solar system by using Eq.
(\ref{oo}).  Notice that ${\bm l}(t)=\lim_{s_g\rightarrow t}{\bm
l}(s_g)\equiv{\bm p}$, and $\theta(t)=\lim_{s_g\rightarrow
t}\theta(s_g)\equiv\Theta$.  The argument of the logarithm in Eq.
(\ref{3a}) for $\varphi_{\rm cal}$ is
\begin{equation}
\label{ss}
-k_\alpha \rho^\alpha=\frac{\nu\rho}{c}\Bigl[1-\cos\theta(s_g)\Bigr]\;.
\end{equation}
The post-Newtonian expansion of the model function of the
electromagnetic phase yields
\begin{equation}
\label{ww}
\varphi_{\rm cal}=\varphi_0+k_\alpha
x^\alpha+\frac{2GM_J}{c^2}\left(k_\alpha
u^\alpha\right)\ln\Bigl\{1-{\bm k}\cdot\bigl[{\bm p}+\frac{1}{c_g}{\bm
p}\times({\bm v}_J\times{\bm
p})\bigr]\Bigr\}+O\left(\frac{v^2_J}{c_g^2}\right),
\end{equation}
where the constant and $\ln\rho$ terms
have been neglected because they are smaller than the other
relativistic terms, and are much smaller than the experimental
angular resolution.

The term in square brackets of the logarithmic function describes a
small displacement of vector ${\bm p}$ to the retarded direction
\begin{equation}
\label{da}
{\bm l}={\bm p}+\frac{1}{c_g}\,{\bm p}\times({\bm v}_J\times{\bm
p})\;.
\end{equation}
This equation is independent of the electromagnetic wave vector ${\bm
k}$ and describes a change from the present (ephemeris) position of
Jupiter given by vector ${\bm p}$ to its retarded location given
by vector ${\bm l}$ (see figure \ref{vlbi}). This displacement is
caused by the propagation of gravity force in the near zone of Jupiter with finite speed, and is
associated with the near-zone aberration of Jupiter's gravity field.

The determination of $\varphi_{0}$, the unperturbed direction to the
quasar ${\bm k}$, and coordinates of Jupiter in $\varphi_{cal}$ can be
obtained to sufficient accuracy using the JPL ephemeris of the solar
system objects with respect to the quasi-inertial reference frame,
defined by the International Celestial Reference Frame grid of quasars
in the sky \cite{fk-apj}.  This allows the measurement of the
retardation of gravity incorporated in Eq. (\ref{1a}) for
retarded time $s$.

As we emphasized in previous sections, the aberration of gravity (\ref{da}) is
unobservable in the case of the gravitational interaction between two slowly-moving
bodies \cite{carl}.  However, the gravitational perturbation of the electromagnetic phase (not the acceleration)
of the radio wave was measured in the experiment in order to extract
information about the direction to the position of Jupiter which, in
addition to the well-known (radial) Einstein light deflection, has
gravitational drag of the radio wave (tangential deflection) by means of the
gravito-magnetic force produced by its orbital motion.  This gravito-magnetic dragging is equivalent to the aberration of the gravity force.

At the time of the closest approach of Jupiter to the quasar on 2002
September 8, the angle $\Theta\sim 3.7'$ $=10^{-3}$ rad, so that
$1-{\bm k}\cdot{\bm p}=(1/2)\Theta^2+O(\Theta^4)$. On the other hand the post-Newtonian parameter $v_J/c = 4.5\times 10^{-5}\ll\Theta$. Hence $v_J/(c\Theta)\ll 1$, and
the aberration of gravity post-Newtonian
Eq. (\ref{ww}) can be expanded \footnote{Notice that this expansion is valid until the relationship $v/c < \Theta$ is hold. It is incompetent to take the limit $\Theta\rightarrow 0$ for fixed value of $v/c$ as mistakenly done by Asada \cite{ass-ijmp}} with respect to a small parameter
$v_J/(c\Theta)$. It yields equation
\begin{equation}
\label{ww1}
\varphi_{\rm cal}=\varphi_0+k_\alpha
x^\alpha+\frac{4GM_J}{c^2}\left(k_\alpha
u^\alpha\right)\left[\ln\Theta-\frac{{\bm k}\cdot{\bm v}_J-({\bm
k}\cdot{\bm p})({\bm p}\cdot{\bm
v}_J)}{c_g\Theta^2}\right]+O\left(\frac{v^2_J}{c_g^2}\right)\;,
\end{equation}
which was used to estimate the magnitude of the aberration of gravity
effect \cite{fk-apj} described by the second term in square
brackets. The magnitude of the aberration of gravity is inversely
proportional to the speed-of-gravity parameter $c_g$ which has been
measured in this experiment as $c_g=1.06\pm 0.21$ in geometrized units
\cite{fk-apj}. Higher angular accuracy is needed in order to detect TT
gravitational waves and measure their speed directly in the wave zone with existing
\cite{ligo} or future \cite{lisa} technologies of the
gravitational-wave detectors and/or other technique \cite{brag}.

\section{Common Misconceptions Associated with This Experiment}

\subsection{Misconception 1: The gravitational field of a uniformly
moving body is static}

This misconception is associated with the following reasoning.  The
gravitational field of a uniformly moving body can be obtained by
solving Einstein's equations in the static frame of the body with its
subsequent Lorentz transformation to the moving frame. Since the
gravitational field in the body's reference frame is static and not
propagating, this ``non-propagation" property of the gravitational
field will be preserved after making the Lorentz transformation to the
moving frame (see, for example, \cite{will-homepage}).

Any field is called propagating if it is obtained as a solution of the
wave-type equation. A field is static if, and only if, it is obtained
as a solution of elliptic-type equation. But the linearized Einstein's equations in
the moving frame are not of the elliptic-type, they contain time
derivatives and are essentially of the wave
Eqs. (\ref{gfe}). The causal (retarded) solution of these
equations describes the gravitational field of a uniformly moving body
which depends on time.  This dependence is represented in the form of
a causal gravity wave, Eq. (\ref{1}), propagating from the moving body (Jupiter) to the field
point $x^\alpha$ along the hypersurface of gravity null cone
(\ref{1a}). The gravity null cone structure is hidden in the static
solution of Einstein's equations in the rest frame of the body.  But,
a Lorentz transformation from the static frame of the body to a moving
frame preserves the structure of the gravity null cone, as Einstein's
equations insist, but now makes the cone characteristics measurable by observing the retarded position from which the gravity field of the moving body deflects light (see Figs. 1--3).
As discussed previously, the null-cone gravity-field characteristics become
measurable in the moving frame because the time derivatives of the
Einstein equation are normalized to the fundamental speed (of gravity), and these terms produce the retarded gravitational field, which is not static.

\subsection{Misconception 2: The propagation of the gravitational field of a
uniformly moving body can not be observed}

The gravitational force acting on a test particle moving in the field
of a uniformly-moving massive body is directed in the first approximation to the present position
of the body in accordance with Eqs. (\ref{j1}) and (\ref{poq}).
Thus, it is assumed that the propagating property (retardation) of the
gravity field of a uniformly moving body cannot be observed \cite{will-homepage}.

This is true if the trajectory of the test particle is the only
measurable quantity.  The normal use of relativistic celestial
mechanics generally deals with such problems as the motion of
binary systems and solar system objects which velocities are much smaller than the fundamental speed of gravity.  However, this is not true
if the phase of an electromagnetic wave is measured as its
front propagates in the field of a uniformly moving body. As GR predicts and this is
shown in section \ref{rp}, the phase depends on the retarded position
of the body.  This retarded position is caused by the 
finite value of the fundamental speed of gravity in the gravity null cone Eq. (\ref{1a}); hence
the fundamental speed for gravity can be measured (see appendix for further details).

\subsection{Misconception 3: The parameter $c$ of the Lorentz transformation is physically the speed of light}

The parameter $c$ of the Lorentz transformation is generally
associated with the speed of light \cite{ass,will-apj,samuel}. However, verbatim interpretation
of $c$ as a physical speed of light narrows the true meaning
of the Lorentz transformation as describing the transformation
property of Maxwell's equations only. Einstein's equations are also
wave equations where the wave operator depends on the fundamental
constant $c$ which must be interpreted as the speed of propagation
of gravitational field. Consequently, the parameter $c$ in the Lorentz transformation
associated with the Einstein gravity field equations must be related to the speed of
gravity. Generally speaking, parameter $c$ of the Lorentz
transformation is just a fundamental constant of the Minkowski space which characterizes an
ultimate speed of propagation of any massless field in vacuum. Hence, the name
of the ``speed of light" for the parameter $c$ in the Lorentz
transformation should not be taken literally when one
interprets gravitational experiments dealing with non-stationary
(propagating) gravitational fields.

\subsection{Misconception 4: The speed of gravity can only be measured
through observation of gravitational waves}

The prevailing view is that the speed of gravity can only be
associated with the propagation of plane gravitational waves decoupled
from the system emitting these waves \cite{will-homepage}.  Such waves are generated if the
second time derivative of the quadrupole moment of the system is not
constant \cite{mtw,ll}; that is, the bodies comprising the system
accelerate.  These gravitational waves, which are type N
according to Petrov's classification of gravitational fields
\cite{pir, frol, petr}, are indeed generated if, and only if, the
bodies accelerate.  However, the gravitational field of a system
emitting gravitational waves also has contributions from other
Petrov's type fields -- I, II, and III, which have in general
relativity the same structure of the null characteristics as the
N-type waves, but decay faster with distance from the system so that
these near-zone fields can not be observed as freely propagating waves of type N
\cite{pir}. Hence, any measurement of the null characteristics of
the near-zone gravitational field provides us with the upper limit on the speed
of propagation of gravitational waves of type N.

Propagation of light through the multipolar gravitational field of an isolated
astronomical system, emitting gravitational waves, in the near,
intermediate, and wave zones of a system, have been already studied
\cite{ksge,kop-pol,kop-kor}.  We showed that parameters of the light signal (wave vector, polarization, etc.) depends in general on the retarded value of the multipole
moments of the system, as well as their first (velocity associated) and
second (acceleration associated) time derivatives. The second time
derivatives of the multipolar field dominate in the wave zone but
plays no significant role in the near and intermediate zones where
the non-radiative gravitational fields associated with the retarded value of the
multipole moment of the system and its first derivatives prevail. This fields and their propagators can be studied in the light-ray deflection experiments when light moves in the near zone of the isolated system.
Thus, a propagating gravitational field should not be associated only
with the accelerated motion of the bodies when measurements are made
in the near and/or intermediate zone of the system emitting
gravitational waves.

   In summary, the propagation of gravity is a general concept which
must be associated with the wave nature of gravitational field, and is
not limited exclusively to plane gravitational waves of Petrov's N
type. Thus, a uniformly moving body generates a retarded gravitational
field which is associated with velocity of the body.  This
retarded velocity-field can be also obtained by making Lorentz
transformation of the static gravitational field. This transformation
induces the aberration of the null characteristics of the gravity
field which can be observed using the phase of the radio waves and
used to measure their speed without direct detection of gravitational
waves, as explained in previous sections of this paper.

The velocity-field is well-known in electrodynamics \cite{jack, ll,
gri} when electromagnetic field of moving charge is calculated in
terms of the retarded Li\'enard-Wiechert electromagnetic potentials. This
velocity-field reflects the propagation of the electromagnetic field
of a uniformly moving charge but it can be also calculated by making
use of the Lorentz transformation. The same is true for gravity. 

\subsection{Misconception 5: The retarded position of Jupiter is taken at the time of the light-ray closest approach}

The consensus model used to evaluate the deflection of light by
Jupiter (egs.~\cite{modest}) calculates the time $t^*$ of closest approach
of the unperturbed light ray to Jupiter, determines the `retarded'
position of Jupiter ${\bm x}_J(t^*)$ at this time, and then calculates the deflection
magnitude and direction.  This interpretation of the retarded position of the light-ray deflecting body (Jupiter) sounds very plausible
but the mathematical solution of the Einstein and eikonal
equations reveals that the retarded position of Jupiter must be
actually taken at the point ${\bm x}_J(s)$ on Jupiter's world line
at the retarded time $s$ defined by the solution of the gravity null cone Eq. (\ref{1a}).

General relativity unequivocally predicts that the deflection of the
radio photon occurs the most strongly when Jupiter is at its retarded
position ${\bm x}_J(s)$ on the gravity null cone (\ref{1a}).  
In the case of a small impact parameter of the light ray to
Jupiter, the time of the closest approach $t^*$ of the light ray to
Jupiter is nearly equal to the retarded time $s$, Eq. (\ref{1a}),
caused by the finite speed of gravity. Thus, the retarded position
${\bm x}_J(s)$ of Jupiter can be expanded around the time of the
closest approach,
${\bm x}_J(s))={\bm x}_J(t^*)+(s-t^*){\bm v}_J(t^*)+...$ so that
position of Jupiter ${\bm x}_J(t^*)$ taken at the time of the closest
approach can be used in calculation of the light-ray deflection with a
very good approximation.  One should not however confuse the cause and
effect -- the time $t^*$ of the closest approach of the photon to Jupiter is
not inherent to the theory at all, but comes from an approximation of
small impact parameter of light-ray trajectory to the light-ray
deflecting body.  It does not mean that the retardation of the gravitational perturbation of light is caused by
light propagation.  Full treatment of the small and large impact
parameter approximations in the case of gravitational lensing is given
in \cite{ks,ksge,km,kop-kor}.

\subsection{Misconception 6: Null characteristics of gravity field can be observed
only at $(v/c)^2$ terms beyond the Shapiro delay}

This misconception arises from the post-Newtonian expansion of the
metric tensor perturbation $h_{\alpha\beta}$ in Eq. (\ref{1})
made prior to solving the eikonal Eq. (\ref{eik}). Indeed, the
post-Newtonian expansion of $h_{\alpha\beta}$, using the Taylor
expansion of the retarded variables depending on the retarded time $s$
in Eq. (\ref{1a}) around the present time $t$, leads to
\begin{equation}
\label{era}
h_{\alpha\beta}(s,{\bm x})=h_{\alpha\beta}(t,{\bm x})+\mbox{terms of order $(v/c)^2$}\;,
\end{equation}
where $h_{\alpha\beta}(t,{\bm x})$ are
solutions of the elliptic-type equations (Newtonian
potentials). Eq. (\ref{era}) suggests that the retardation of
gravity effect is postponed to the terms of order $(v/c)^2$ because no
terms of order $v/c$ is {\it explicitly} present in this expansion \cite{will-apj,carlip}.

The physical reason for the absence of terms of order $v/c$ in
Eq. (\ref{era}) is that the Li\'enard-Wiechert potentials depend
on distance $r_R=-u_\alpha r^\alpha$ which incorporates both the
retardation and aberration of gravity effects that are equivalent in
terms of order $v/c$ and cancel each other when the post-Newtonian
expansion of the metric tensor is done. Light propagates in GR with the same speed as gravity. Lorentz-invariant solution of the eikonal
equation incorporates the aberration of light which is equal to the aberration of gravity terms and mutually cancel each other, hence, leaving the
retardation of gravity terms in the radio wave phase unchanged. That
is why the post-Newtonian expansion of the retarded gravitational perturbation of the electromagnetic phase
reproduces the aberration of gravity terms which we had observed in
the Jovian deflection experiment via the retarded position of Jupiter
\cite{fk-apj}.

Although $h_{\alpha\beta}(t,{\bm x})$ does not contain terms of order
$v/c$, the measurement of these Newtonian-like gravitational potentials by VLBI is
indeterminate because the gauge invariance of the Einstein equations
make the potentials coordinate-system dependent.  There are gauges,
like the ADM gauge \cite{mtw}, where $g_{00}$ and $g_{0i}$ components
of the metric tensor become the solution of the elliptic-type
equations which suggest that these metric tensor components do not
propagate at all. But, of course, any gravitational influence of these
potentials on measurable quantities is not instantaneous but
propagates with the finite speed. This resembles the situation in
electrodynamics where the Coulomb gauge leads to the elliptic-type
equation for scalar potential but it does not mean that electric and
magnetic fields propagate instantaneously \cite{gri}. Hence, the
potentials are not directly measurable quantities in the light-ray deflection experiments because they depend
on the specific choice of gauge conditions which is a matter of
convenience.  Only those observable quantities which are
gauge-independent (like the electromagnetic phase) can be used to probe physical properties of the
gravitational fields.

Our method of analysis, given in this paper, consists of two part: first, to
solve the problem of the propagation of light in the time-dependent
gravitational field by making use of the Li\'enard-Wiechert
potentials; and then to determine the post-Newtonian expansion of the
solution in order to produce the physical interpretation in terms of
different order in $v/c$.  However, we will show below that if the
post-Newtonian potentials (\ref{era}) are used to solve the eikonal
Eq. (\ref{eik}), the solution will contain terms of order $v/c$
which can be represented in several equivalent forms---leading to
difficulty in the interpretation of the solution. This difficulty can be
resolved if the solution of the eikonal equation at higher orders
of $v/c$ is also determined in order to match it smoothly with the linear terms of order $v/c$ present in the phase.

   For this experiment, the observable quantity is electromagnetic
phase $\varphi$ which is a gauge-independent quantity \cite{mtw}. We
can measure the impact of the gravitational potentials on the
electromagnetic phase in the following manner. The post-Newtonian
expansion of the electromagnetic phase is
\begin{equation}
\label{zx}
\varphi(x^\alpha)=\varphi_0+k_\alpha x^\alpha+\frac{1}{c^2}\psi(x^\alpha)\;,
\end{equation}
where $\varphi_0$ is constant and $k^\alpha$ is the wave vector of
electromagnetic phase. Substitution of Eq. (\ref{zx}) to the
eikonal equation and making use of expansion (\ref{era}) yields the linearized equation for the phase perturbation $\psi$
\begin{equation}
\label{zo}
\frac{1}{c}\frac{\partial\psi}{\partial t}+k^i\frac{\partial\psi}{\partial x^i}=2\left(1-\frac{{\bm k}\cdot{\bm v}_J}{c}\right)U_J(t,{\bm x})+O\left(\frac{v_J^2}{c^2}\right)\;,
\end{equation}
where $U_J(t,{\bm x})=GM_J/R$, $R=|{\bm x}-{\bm x}_J(t)|$, and $k^i$
is the unit vector in the direction of propagation of light rays which
are orthogonal to the hypersurface of the phase $\varphi$.  Eq.
(\ref{zo}) is a partial differential equation of the first order and
its solution is
\begin{equation}
\label{xq}
\psi(x^\alpha)=2GM_J\biggl(1-\frac{{\bm k}\cdot{\bm v}_J}{c}\biggr)\ln\biggl[pR-{\bm p}\cdot{\bm R}\biggr]+O\left(\frac{v_J^2}{c^2}\right) \;,
\end{equation}
where ${\bm p}={\bm k}-c^{-1}{\bm v}_J$, and ${\bm v}_J=d{\bm x}_J(t)/dt$ is Jupiter's velocity. 

The argument of the logarithmic function in Eq. (\ref{xq}) can be
represented in two different forms. The first form is given by
\cite{will-apj, kop-pla}
\begin{eqnarray}
\label{h1}
pR-{\bm p}\cdot{\bm R}=\biggl(1-\frac{1}{c}{\bm k}\cdot{\bm v}\biggr)\biggl(R-\bm{\mathcal{K}}\cdot{\bm R}\biggr)\;,
\end{eqnarray}
where
\begin{eqnarray}
\label{h2}
\bm{\mathcal{K}}\equiv\frac{\bm p}{p}={\bm k}-\frac{1}{c}\Bigl({\bm k}\times({\bm v}_J\times{\bm k}) \Bigr)\;.
\end{eqnarray}
The second form is 
\begin{equation}
\label{h3}
pR-{\bm p}\cdot{\bm R}=r-{\bm k}\cdot{\bm r} = -k_\alpha r^\alpha\;,
\end{equation}
where $k^\alpha$ is a wave vector of the radio wave, $r^\alpha=(r,
{\bm r})$, ${\bm r}={\bm x}-{\bm x}_J(s)$, $r=|{\bm r}(s)|$, and the
retarded time $s$ is given by the gravity null cone Eq.
(\ref{1a}).

Eq. (\ref{h2}) looks similar in form to the aberration of light
formula transforming the direction of propagation of the light ray ${\bm
k}$ from the moving frame of an observer to the direction
$\bm{\mathcal{K}}$ in the static frame of Jupiter. This is why Will
\cite{will-apj} and Carlip \cite{carlip} interpreted the effect we have observed as the
aberration of light (see next section and appendix).  However, Eq. (\ref{h3}) clearly
shows that this aberration term should be interpreted as an aberration
of gravity caused by its propagation because Jupiter's coordinate
depend on the retarded time $s$ defined through the solution of the
Einstein (wave) Eq. (\ref{gfe}).  Any aberration, of course, is
of order $v/c$ but only one type of aberration of either
electromagnetic or gravitational field can be unambiguously extended
to terms of second order in $v/c$ to keep Lorentz-invariance of the
equations. Our calculations show that aberration term in Eq. (\ref{h2}) matches smoothly with the retardation of gravity effect (\ref{h3}) in any post-Newtonian order while the aberration of light does not (see also \cite{kl} for more detail on the matching procedure).

\subsection{Misconception 7: The Jovian experiment and the PPN parameter $\alpha_1$.}

The Jovian deflection experiment measures the aberration of
gravity and its fundamental speed limit.  It tests for the first
time the non-stationary gravito-magnetic property of the gravitational
field, of order $(v/c)^3$, generated by the time derivative of the metric tensor (gravito-magnetic displacement
current).  The experiment can also be viewed as a test of the
Lorentz-invariance of the gravitational field and confirmation of its relativistic causal nature. 

Will \cite{w-book} introduced the ``preferred frame" PPN parameter
$\alpha_1$ which characterizes the violation of the Lorentz-invariance
of the PPN metric tensor. However, this parameter presents only in $g_{0i}$ component of the metric tensor if it is written down in the preferred frame. Hence, parameter $\alpha_1$ characterizes, in fact, a strength of the physical coupling between the $g_{0i}$ component of the metric tensor and the matter current given by $T_{0i}$ component of the stress-energy tensor. Subsequent Lorentz transformation to a frame moving with respect to the preferred frame with velocity $\vec{w}$ introduces a product $\alpha_1\vec{w}$ to $g_{00}$ and $g_{0i}$ components of the metric tensor.

Under the assumption that the cosmic
microwave background radiation (CMBR) defines the preferred frame,
several accurate experimental tests give $\alpha_1\le 4\times
10^{-4}$, which appears more accurate than our limit on $(c_g/c)-1\le
0.27$, both of which are measures of the experimental departure from
GR.  However, direct comparison between these two methods for testing
Lorentz-invariance is not valid.  First, the PPN formulation states that $\alpha_1$ is not coupled with time derivatives of the metric tensor whereas the
deflection experiment is a direct test of the magnitude of the time derivative of the metric tensor (see Eq. (\ref{dfg}) and \cite{kop-cqg,kop-ijmp1}). In GR the parameter
$\alpha_1\equiv 0$. Therefore, search for
$\alpha_1$ is an attempt to find an effect, which probably does not exist at all because GR is a valid theory, while our
measurement of the fundamental speed of gravity discovers a real (and newly measured)
physical effect -- the aberration of gravity, which is a fundamental property of gravity predicting that electromagnetic signals are deflected by gravitational field of a moving gravitational body from its retarded position with respect to observer. 

Secondly, the current limit on the parameter $\alpha_1$ strongly depends on
our guess about the velocity $\vec{w}$ of the solar system relative to a ``preferred
frame" which was assumed to be the CMBR.  But, we do not know if it
defines the preferred coordinate system in the sense
forbidden by special relativity; it is just a convenient frame to
describe the global isotropy of the cosmological black-body radiation. Furthermore, future observations of the
relic gravitational wave background (GWB), for example, may lead to
the frame somehow moving with respect to the CMBR frame. Since the GWB was
formed in the very early universe, long before the CMBR decoupled from
matter, it would be more likely to associate it with the preferred frame.
However, neither the global topological isotropy of CMBR nor GWB may be related to the local Lorentz-invariant isotropy of
the space-time itself.  Finally, modern multi-connected cosmologies
challenge the cosmological Copernican Principle, and indicate that
various possibilities for the global preferred frame may not coincide with
the CMBR and GWB frames \citep{blev}.  Thus, the limit on $\alpha_1$,
established by \cite{w-book} depends on numerous assumptions about properties of a preferred
reference frame and other cosmological principles which are to some extent ambiguous concepts. It seems to us that only GP-B experiment will be able to set up a limit on $\alpha_1$ parameter which will not be sensitive to the choice of the preferred frame \cite{gpb}.

\section{Other Interpretations}

Four alternative interpretations, which disagree with our
interpretation and among themselves, of the Jovian deflection
experiment have been proposed (see section \ref{intro}).  We have analyzed fully the alternative
interpretations elsewhere \cite{kop-cqg,kop-ijmp1,kop-ijmp2,kop-pla2}, but summarize the main
arguments here.

\begin{figure*}
\includegraphics[keepaspectratio=true, width=16cm,height=16cm]{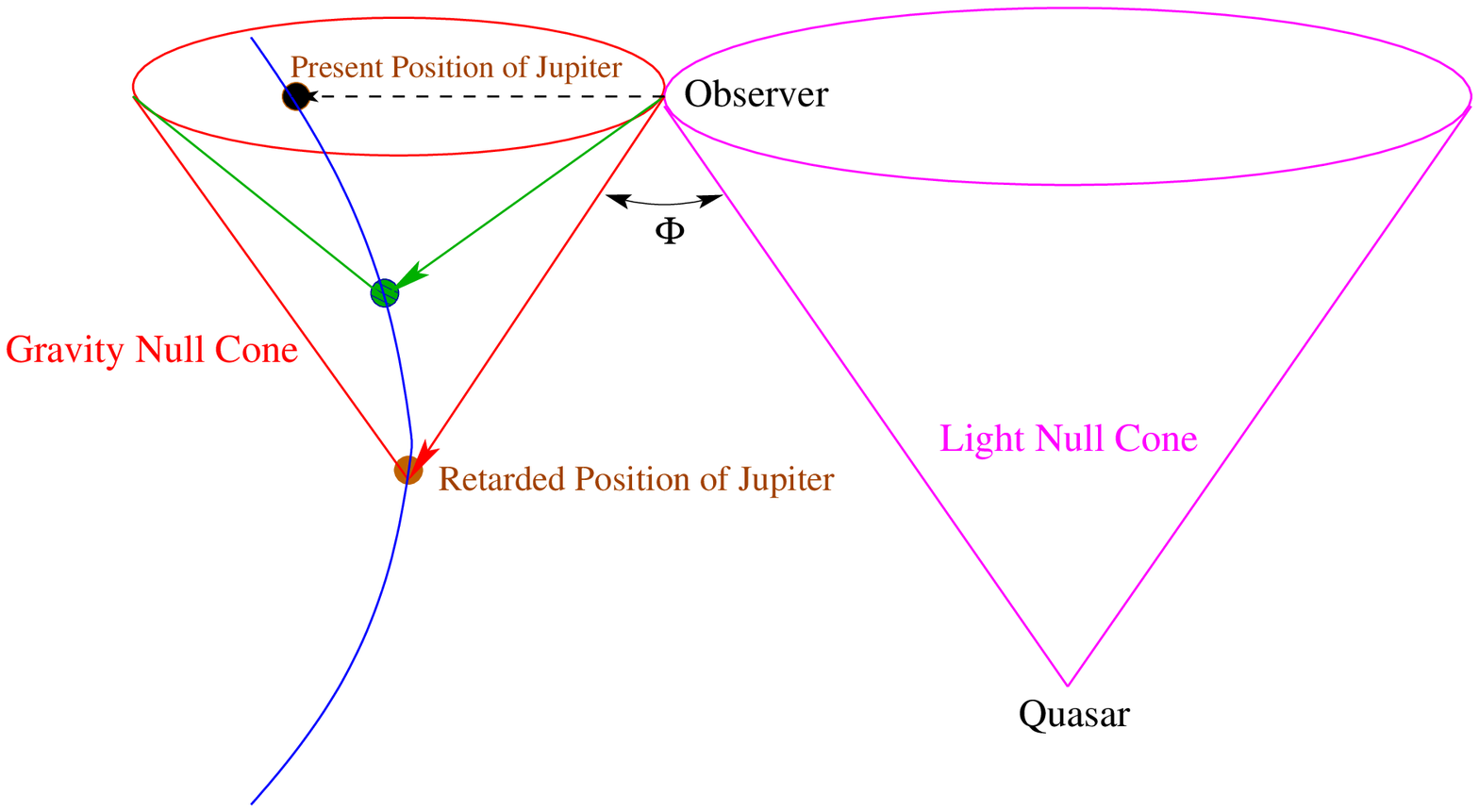}
\caption{\label{null}The two null cones related to the experiment are
shown. The gravity cone is a solution of the Einstein equations and
describes null characteristics of gravity originating from moving Jupiter. The light cone is a
solution of the Maxwell equations and describes the null characteristics of a radio
wave from a quasar. The gravitational deflection of light is seen at the point where the gravity cone of Jupiter passes
through the observer. VLBI measures the Minkowski dot
product $\Phi=-k_\alpha r^\alpha$ between two null vectors $k^\alpha$
and $r^\alpha=x^\alpha-x^\alpha_J(s)$ at the point of observation
directed to the observer from the quasar and from Jupiter,
respectively. Had moving Jupiter not been detected at the retarded position
on its world line, the speed of gravity would not be equal
to the speed of light and the general theory of relativity would
be violated.}
\end{figure*}

\begin{figure*}
\includegraphics[keepaspectratio=true,width=17cm,height=17cm]{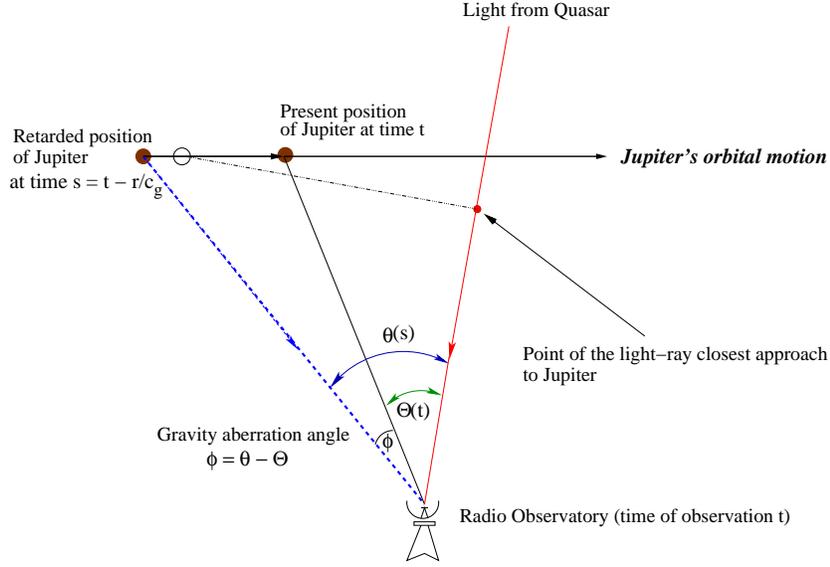}
\caption{\label{vlbi} Light propagates from the quasar toward a VLBI
station (observer) on the Earth. While the light propagates, Jupiter
is moving. General relativity predicts \cite{kop2001, fk-apj} that the
light observed at the time $t$ is deflected most strongly by Jupiter
when it is located at the retarded position ${\bm x}_J(s_g)$
($s_g=t-r/c_g$) for $c_g=c$, regardless of the direction of propagation of the light
ray and the magnitude of the light-ray impact parameter with respect
to Jupiter. This property of the gravitational deflection of light by a moving massive body can be used in order to measure the aberration of gravity force with respect to the aberration of light. The region of the sky which is imaged by the VLBI
observations is smaller than the angle $\Theta$, so that radio
emission of Jupiter is not detected when the quasar is observed. }
\end{figure*}

The main reason for the disagreements between the four other
interpretations is their use of the truncated linearized expansion of
the Einstein equations which is not capable to study the Lorentz-invariance of the gravity field force in the light-ray deflection experiments in higher orders in $v/c$.  The
truncated linearized theory is not sufficient to give the proper
geometrical picture of the observed quantity in the time delay of light which is, in fact, a
Minkowskian dot product $\Phi=-k_\alpha r^\alpha$ between two null
vectors $k^\alpha$ and $r^\alpha$ (see figure \ref{null}) entering the
gravitational perturbation of the electromagnetic phase $\varphi$ in
Eq. (\ref{3}). When Einstein's theory of the gravitational time delay is truncated at linear $v/c$
order terms, the simple geometric picture of the experiment is easily
missed and the interpretation of the experiment becomes obscured.

Will \cite{will-apj} (see also \cite{carlip}) introduced a set of alternative theories of
gravity with a propagation speed $c_g$ which could differ from the
speed of light $c$.  However, the use of the Post-Newtonian
Parameterized (PPN) approximation, which is wide-spread, does not keep
the geometrical relationship between the Christoffel symbols (light
geodesic equations) and Ricci tensor for different values of
$c_g$. Moreover, the PPN formalism transforms the Christoffel symbols from one inertial frame to another with the matrix of the Lorentz transformation for electromagnetic field \cite{kop-ijmp1}. This makes no sense since the Christoffel symbols are defined purely in terms of gravitational potentials which must be transformed in accordance with the group of the Lorentz transformation of gravitational field depending on the parameter $c_g$. In addition, the PPN gravity field equations with $c_g$ are not derivable from a Lagrangian of any known theory of gravity. Hence, they do not provide us with a complete physical description of the gravitational field in case when the speed of gravity $c_g\not=c$ making the PPN approach unreliable for analysis of experiments in time-dependent gravitational fields.  
This may explain why Will's use of the PPN approximation to analyze relativistic
corrections to the Shapiro time delay confuses the aberration of light and the aberration of gravity (see appendix for more detail).

   As discussed in the previous section, Will's formulation does
predict an `aberration' deflection of order $v/c$, Eq.
(\ref{h2}), although Will attributes this aberration to propagation of ``light".  The
difference between his interpretation with ours can be resolved by
finding explicit terms of the second order in $v_J/c$ in the solution
of eikonal Eq. (\ref{eik}), and then matching them smoothly with
the solution (\ref{xq}).  This matching shows \cite{kop-cqg} that the
correct interpretation of the retarded position of Jupiter in the gravitational perturbation of electromagnetic phase in Eq.
(\ref{xq}) must be associated with the null characteristics of gravity field.  In other words,
had the exact aberration-of-light formula been used to account for the
second order terms in Eq. (\ref{h2}), it would not produce {\it
all} of the second and higher order terms in the solution of the
eikonal Eq. (\ref{eik}), whereas the retarded time Eq.
(\ref{h3}) gives all terms of second and higher orders in the Lorentz
covariant solution of the eikonal Eq. (\ref{eik}) as shown by the
calculations given in the present paper and in \cite{kl,kop-ijmp1,kop-pla2}. Thus, the correct meaning of
the linearized Eq. (\ref{h2}) should not be associated with the
aberration of light but with the retardation (aberration) of gravity caused by the orbital motion of Jupiter
\cite{kop-pla} and associated with the finite value of the fundamental speed of gravity \footnote{AS we mentioned the effect we have observed can be also explained in terms of the gravito-magnetic dragging of the quasar light ray \cite{kop-cqg,kop-ijmp1}}.

Asada \cite{ass} suggested that the origin of the retarded time
Eq. (\ref{1a}) is associated with the propagation of the radio
waves from the quasar in the field of static Jupiter.  However, this equation originates
from the Li\'enard-Wiechert solution of the Einstein equations (\ref{gfe}), which describes the
propagation of gravitational field and is not associated in any way with the
propagation of radio waves from the quasar. Propagation of radio
waves is described by the time delay Eq. (\ref{tde}) which
describes the light cone, while the retarded time
Eq. (\ref{1a}) defines characteristics of the gravity field
null cone (see figure \ref{null}). Calculations displayed in the appendix of this paper obviously demonstrate that Asada's interpretation \cite{ass} is illicit since it confuses the null characteristics of light and gravity.

A fundamental flaw in Samuel's \cite{samuel} interpretation was his
assumption that the direction to Jupiter was directly measured by VLBI
network in the deflection experiment so he also confused the null directions associated with characteristics of
gravity and radio waves \cite{kop-ijmp2}.  Even at the minimal
separation of $3.7'$ of Jupiter and the quasar on 2002 September 8, no
radio emission from Jupiter was obtained with the VLBI observations.
The null direction to retarded position of Jupiter, depending on the value of the fundamental speed of gravity, was measured from the gravitational time delay Eq. (\ref{tde}) and compared with that known extremely
accurately from the JPL solar system ephemeris depending on the speed of light (see appendix and Fig. \ref{twocones}). 
Therefore, minimization of the residual electromagnetic phase with the speed of
gravity taken as a fitting parameter allowed us to measure it with respect to the speed of light. 

Pascual-S\'anchez \cite{pask} stated that the retarded position of Jupiter measured in the experiment is the R{\o}mer delay of light coming from Jupiter to observer and that this  R{\o}mer delay has nothing to do with the gravitational time delay of light. Eq. (\ref{1a}) for retarded time $s$ looks indeed similar to the R{\o}mer delay but this delay is for propagation of the gravitational field from moving Jupiter to the observer at Earth as independently proved in the appendix of the present paper and in \cite{kop-ijmp1}. It does not describe the R{\o}mer delay of light but gravity which takes finite time to propagate from moving Jupiter to observer. Had the R{\o}mer delay of gravity been different from that for light, Jupiter would deflect light not from its retarded but from some other position on its orbit. The experiment disproved this assumption confirming that the R{\o}mer delay of gravity and light are equal with accuracy of 20\% by comparing two positions of Jupiter on its orbit obtained independently from the gravitational deflection of quasar's light and JPL ephemeris.

\section{Summary}
We have developed a Lorentz-invariant general-relativistic theory of
light-deflection experiments associated with moving bodies in the
solar system. The formulation is based on the retarded
Li\'enard-Wiechert solution of the Einstein equations, and explicitly
calculates the effect on the phase of the radio wave from the quasar
caused by the finite speed 
of gravity which is a primary fundamental constant in Einstein's field equations. Our
post-Newtonian expansion of the retarded coordinate of the light-ray
deflecting body shows that the retardation of gravity effect 
after making its post-Newtonian expansion, is present already in
linear aberration-like terms of order $v/c$ beyond the Shapiro
delay; hence, we interpret the $v/c$ term as caused by the aberration
of gravity in the expression for the gravitational force (\ref{j})
exerted by Jupiter on the photons.  We have used the concept of the
gravito-magnetic field caused by translational motion of Jupiter as an aid in understanding the gravitational
phenomena associated with this experiment.

The VLBI measurements confirmed that the deflection is associated with
the retarded position of Jupiter, with the speed of gravity equal to that of light to 20\% accuracy
\cite{fk-apj} and this retardation is not a preferred frame
effect associated with the PPN parameter $\alpha_1$.  Other
contradictory interpretations of the experimental results are
associated with: (1) an analysis of the Einstein equations which are
not fully Lorentz-invariant; (2) misconceptions about the experimental
method and analysis; (3) misunderstanding of the nature of the
interaction of the gravitational field from moving objects with the
{\it phase} of the quasar radio signal which is a null hypersurface. 
According to GR the phase is influenced in {\it all} orders of $v/c$ by the
propagating gravitational field of Jupiter and this perturbation
reveals itself in the linearized order of $v/c$ as the aberration of gravity force of Jupiter due to
the finite speed of gravity.

We thank the Eppley Foundation for Research (award 002672) for
support.

\appendix
\section{Propagation of Light in a Bi-metric Theory}
In this section we assume that the Einstein theory of general relativity is valid with the time coordinate $x^0=ct$ where $c$ denotes the fundamental speed of gravity. Let us assume that light propagates in vacuum with the speed $c_l$ less than the speed of gravity $c$. Effectively, this means that vacuum in the presence of the gravity field can be considered as a fully transparent medium which has a constant refractive index $\epsilon\ge 1$. Propagation of light in such vacuum can be treated on the basis of the Maxwell electrodynamics in medium. The geometric optics limit of these equations has been elaborated by Synge  \cite{synge} and we employ his approach.

According to Synge \cite{synge} light propagates in the medium along light geodesics of an optical metric defined by
\begin{eqnarray}\label{A1}
\bar{g}_{\alpha\beta}&=&{g}_{\alpha\beta}+\left(1-\frac{1}{\epsilon^2}\right)V_\alpha V_\beta\;,\\
\label{A2}
\bar{g}^{\alpha\beta}&=&{g}^{\alpha\beta}-\left(\epsilon^2-1\right)V^\alpha V^\beta\;,
\end{eqnarray}  
where ${g}_{\alpha\beta}$ is the gravitational metric given by Eqs. (\ref{y}), (\ref{1}), and $V^\alpha$ is a vector field associated with four-velocity of the medium. In our case it means that if the speed of gravity and light are not equal there must exist a preferred frame in which the speed of light is isotropic while in all other frames it is anisotropic. Notice that in the preferred frame $V^\alpha=(1,0,0,0)$ and the speed of light $c_l=c/\epsilon < c$.

There are other bi-metric theories of gravity (see, for example, \cite{carlip,jm1}) with two metrics predicting different speeds of propagation for light and gravity. In such theories the four-velocity $V^\alpha$ is identified with a long-range vector field spontaneously violating the Lorentz-invariance of gravity. Those bi-metric theories lead to the geodesic equations of light also propagating along the null cone of the optical metric \cite{carlip}.

Let us assume for simplicity that both the refractive index $\epsilon$ and the vector field $V^\alpha$ are constant \footnote{More general case can be solved as well but it significantly complicates solution of equations and will be omitted}. This model is simple but self-consistent and sufficient to apparently distinguish the speed of light $c_l$ from the speed of gravity $c$. Equations of our bi-metric model and their solution are given in the rest of this appendix.

Light rays in Synge's approach \cite{synge} are defined by a covariant equation for electromagnetic phase (eikonal) $\varphi$ which reads \footnote{Lagrangian-based approach of an alternative bi-metric theory of gravity \cite{carlip} leads to the same equation.}
\begin{equation}
\label{A3}
\bar{g}^{\mu\nu}\partial_\mu\varphi\partial_\nu\varphi=0\;,
\end{equation}
which generalizes Eq. (\ref{eik}) for propagation of light in the case of $c_l\not=c$. Assuming that unperturbed solution of Eq. (\ref{A3}) is a plane wave we can write a general solution of this equations as follows
\begin{equation}
\label{A4}
\varphi(x^\alpha)=\varphi_0+k_\alpha x^\alpha+\psi(x^\alpha)\;,
\end{equation}
where $k_\alpha$ is an unperturbed (constant) wave co-vector of the electromagnetic wave, and $\psi(x)$ is a relativistic perturbation of the eikonal generated by the metric tensor perturbation ${h}_{\alpha\beta}$ defined in Eq. (\ref{1}). Substitution of Eq. (\ref{A4}) to (\ref{A3}) yields
\begin{eqnarray}\label{A5}
\bar{\eta}^{\alpha\beta}k_\alpha k_\beta&=&0\;,\\
\label{A6}
\bar{\eta}^{\alpha\beta}k_\alpha\frac{\partial\psi}{\partial x^\beta}&=&\frac{1}{2}{h}^{\alpha\beta}k_\alpha k_\beta\;,
\end{eqnarray}
where the unperturbed part of the optical metric (\ref{A1}), (\ref{A2}) is defined by
\begin{eqnarray}
\label{A7}
\bar{\eta}^{\alpha\beta}&=&{\eta}^{\alpha\beta}-\left(\epsilon^2-1\right)V^\alpha V^\beta\;,\\
\label{A8}
\bar{\eta}_{\alpha\beta}&=&{\eta}_{\alpha\beta}+\left(1-\frac{1}{\epsilon^2}\right)V_\alpha V_\beta\;.
\end{eqnarray}

\begin{figure*}
\includegraphics[keepaspectratio=true,width=15cm,height=15cm]{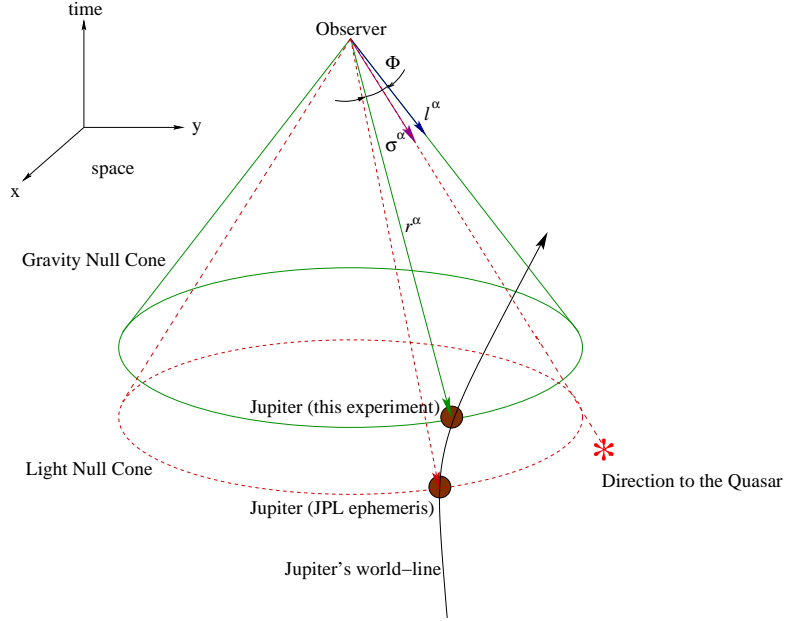}
\caption{\label{twocones}
The two null cones of the bi-metric theory are shown (see appendix). Retarded position of Jupiter due to the finite speed of gravity is on the gravity null-cone while Jupiter's radio position is on the light null-cone. The unperturbed direction to the quasar is defined by vector $\sigma^\alpha$ lying on the light null-cone of the optical metric $\bar{g}_{\alpha\beta}$. Gravity perturbs the light null-cone vector $\sigma^\alpha$ and changes its direction to the gravity null-cone vector $l^\alpha$. Gravitational field of moving Jupiter deflects the quasar's light from the retarded position defined with respect to observer by vector $r^\alpha=x^\alpha-x^\alpha_J(s)$ which lies on the gravity null-cone. Perturbed eikonal $\psi\simeq 2M_J\ln\Phi$, where $\Phi=-l_\alpha r^\alpha$. Components of the vector $l^\alpha$ are calculated from vector $\sigma^\alpha$ and four-velocity of Jupiter $u^\alpha$ which are known. The angle $\Phi$ is calculated assuming that the retarded position of Jupiter is defined by the JPL ephemerides and lies on the light null-cone. The observed value of the angle $\Phi$ is fit to that defined by the JPL ephemerides by adjusting the parameter $\epsilon$ defining the difference between the speed of light and gravity. This procedure measures the speed of gravity with respect to light within the accuracy of 20\% for 10 $\mu$as astrometric tolerance in observing small angles in the sky with the phase-referencing  VLBI technique.}
\end{figure*}

Let us introduce a light-ray vector $\sigma^\alpha$ defining direction of propagation of the unperturbed light ray. One has \cite{synge}
\begin{equation}
\label{A9}
\sigma^\alpha=\bar{\eta}^{\alpha\beta}k_\beta=k^\alpha-\left(\epsilon^2-1\right)\left(V^\beta k_\beta\right)V^\alpha\;,
\end{equation}
which implies that 
\begin{equation}\label{A10}
\bar\eta_{\alpha\beta}\sigma^\alpha\sigma^\beta=0\;,\qquad\qquad\mbox{and}\qquad\qquad k_\alpha\sigma^\alpha=0\;.
\end{equation}
Vector $\sigma^\alpha$ points to the direction of propagation of a light ray from a radio source (quasar) to observer (see Fig. \ref{twocones}). Making use of vector $\sigma^\alpha$ simplifies Eq. (\ref{A6}) and reduces it to the following form
\begin{equation}\label{A11}
\sigma^\alpha\frac{\partial\psi}{\partial x^\alpha}=\frac{1}{2}{h}_{\alpha\beta}\sigma^\alpha \sigma^\beta\;.
\end{equation}
The unperturbed characteristics of the eikonal Eq. (\ref{A11}) are straight lines (light rays) parametrized by the affine parameter $\lambda$ in such a way that 
\begin{equation}\label{A12}
\frac{d}{d\lambda}=\sigma^\alpha\frac{\partial}{\partial x^\alpha}\;.
\end{equation}
Integration of Eq. (\ref{A12}) by making use of the unperturbed characteristics is straightforward (see, for example, \cite{zm}) and is given as follows
\begin{equation}\label{Api}
\psi(x^\alpha)=-\frac{2GM}{c^2}\frac{\left(\sigma_\alpha u^\alpha\right)^2+(1/2)\left(\sigma_\alpha \sigma^\alpha\right)}{\left(P_{\alpha\beta}\sigma^\alpha\sigma^\beta\right)^{1/2}}  \ln\Bigl(-{l}_\alpha r^\alpha \Bigr)\;,
\end{equation}
where we have defined 
\begin{eqnarray}\label{A14}
{l}^\alpha&=&\sigma_\bot^\alpha+\sigma_\bot u^\alpha\;,\\
\label{A15}
\sigma_\bot^\alpha&=&P^{\alpha}_{\;\beta}\sigma^\beta\;,\\
\label{A16}
\sigma_\bot&=&\left(\sigma_{\bot\alpha}\sigma^\alpha_\bot\right)^{1/2}=\left(P_{\alpha\beta}\sigma^\alpha\sigma^\beta\right)^{1/2}\;,
\end{eqnarray}
and 
\begin{equation}\label{A17}
P_{\alpha\beta}=\eta_{\alpha\beta}+u_\alpha u_\beta\;,
\end{equation}
is the operator of projection on the hyperplane orthogonal to the four-velocity $u^\alpha$ of the light-ray deflecting body so that it obeys to equation $P_{\alpha\beta}P^\beta_{\;\gamma}=P_{\alpha\gamma}$. It is easy to confirm that Eq. (\ref{Api}) is solution of Eq. (\ref{A11}) by observing that 
\begin{equation}
\label{gimk}
\frac{d}{d\lambda}\ln\Bigl(-l_\alpha r^\alpha\Bigr)=-\frac{\sigma_\bot}{r_R}\;,
\end{equation}
where $r_R=-u_\alpha r^\alpha$, and Eqs. (\ref{qa}), (\ref{po}) have been used.
Eq. (\ref{A14}) allows us to recast the argument of the logarithm in Eq. (\ref{Api}) as
\begin{equation}\label{A18}
{l}_\alpha r^\alpha=\sigma_\bot^\alpha r_\alpha-\sigma_\bot r_R\;.
\end{equation}
It is worth noticing that both vectors ${l}^\alpha$ and $r^\alpha$ are null vectors of the unperturbed gravity metric $\eta_{\alpha\beta}$, that is, 
\begin{eqnarray}\label{A19}
\eta_{\alpha\beta}{l}^\alpha{l}^\beta&=&0\;,\\
\label{A20}
\eta_{\alpha\beta}r^\alpha r^\beta&=&0\;,
\end{eqnarray}
which are consequences of definitions given by Eqs. (\ref{A14}) and (\ref{grav}). Thus, neither $l^\alpha$ nor $r^\alpha$ belong to the null cone of the optical metric $\bar{g}_{\alpha\beta}$ (see Fig. \ref{twocones}). This point has not been understood by researchers whose interpretations of the Jovian deflection experiment differ between themselves. In particular, Asada \cite{ass} insisted that the retarded position of Jupiter is on the null cone of the optical metric. Our calculations show that this Asada's assertion is erroneous.  

Solution (\ref{A4}), (\ref{Api}) for the electromagnetic eikonal in the bi-metric theory should be compared with a similar solution (\ref{3}) for the case of propagation of light in general relativity where gravity and light have the same null cone. The reader can see that the null directions (\ref{1a}) of the gravity metric $g_{\alpha\beta}$  enters explicitly the gravitationally perturbed part of the eikonal (\ref{Api}) in the bi-metric theory in the form of a dot product ${l}_\alpha r^\alpha$ in the argument of the logarithm, where $r^\alpha$ is the null distance of the metric $g_{\alpha\beta}$ between observer and the light-ray deflecting body \footnote{The amplitude of the eikonal also changes leading to the deviation of the PPN parameter $\gamma$ from its general relativistic value $\gamma=1$.}. This term replaces the argument $k_\alpha r^\alpha$ of the logarithmic term in the eikonal (\ref{3}) in general relativity where $c_l=c$. A remarkable fact is that both $l^\alpha$ and $r^\alpha$ are null vectors of the metric $g_{\alpha\beta}$ describing propagation of gravity. Consequently, gravitational light-ray deflection experiments in the field of moving bodies are sensitive to, and can measure, the null characteristics of the gravity metric $g_{\alpha\beta}$ by observing propagation of light that moves along the null characteristics of the optical metric $\bar{g}_{\alpha\beta}$. This fully agrees with our general relativistic interpretation of the Jovian experiment \cite{kop2001,fk-apj}.
  
Analysis given in this appendix provides a firm support to our statement \cite{kop2001,fk-apj} that the retarded coordinate of Jupiter $x^\alpha_J(s)$ measured in the Jovian experiment is associated with the finite speed of gravity but not light. The statement that the Jovian experiment measures the light-cone effects associated with the optical metric $\bar{g}_{\alpha\beta}$ as claimed in \cite{ass,will-apj,carlip,samuel,pask,will-livrev} is not valid.   

\newpage

\newpage
\newpage
\newpage
\end{document}